\documentclass[10pt]{elsarticlehome}
\usepackage{epsfig}

\begin{document}

\title{Emergent gravity and chiral anomaly in Dirac semimetals in the presence of dislocations}


\author[UWO,ITEP]{M.A.~Zubkov
 }



\address[UWO]{The University of Western Ontario, Department of Applied
Mathematics,  1151 Richmond St. N., London (ON), Canada N6A 5B7}

\address[ITEP]{ITEP, B.Cheremushkinskaya 25, Moscow, 117259, Russia
}

\begin{abstract}
We consider the recently discovered Dirac semimetals with two Dirac points $\pm{\bf K}$. In the presence of elastic deformations each fermion propagates in a curved space, whose metric is defined by the expansion of the effective Hamiltonian near the Dirac point. Besides, there is the emergent electromagnetic field that is defined by the shift of the Dirac point. We consider the case, when the deformations are caused by the dislocations. The dislocation carries singular torsion and the quantized flux of emergent magnetic field. Both torsion singularity and emergent magnetic flux may be observed in the scattering of quasiparticles on the dislocation due to Stodolsky and Aharonov - Bohm effects. We discuss quantum anomalies in the quasiparticle currents in the presence of emergent gauge and gravitational fields and the external electromagnetic field. In particular, it is demonstrated, that in the presence of external electric field the quasiparticles/holes are pumped from vacuum along the dislocation. The appeared  chiral imbalance along the dislocation drives the analogue of chiral magnetic effect, that is the appearance of electric current along the dislocation.
\end{abstract}



\maketitle


\newcommand{\revision}[1]{{#1}}
\newcommand{\revisionB}[1]{{#1}}

\section{Introduction}

Condensed matter systems with emergent relativistic invariance may serve as an arena for the experimental observation of various effects typical for the high - energy physics. In particular, some of such effects were predicted and checked expreimentally in $2+1$ D graphene, where the relativistic $2+1$ D fermions appear in the vicinity of the Fermi points (see the detailed review in \cite{Katsbook}). The emergent $3+1$ D relativistic fermions appear in $^3$He-A superfluid. (For the description of various emergent relativistic effects in $^3$He see monograph \cite{Volovik2003}). Recently, the $3+1$ D relativistic fermions were discovered experimentally in the novel material - the Dirac semimetal \cite{semimetal_discovery,semimetal_discovery2,semimetal_discovery3}. For the description of the microscopic theory of Dirac semimetals (that may be approximated by the low energy effective field theory with emergent relativistic invariance) see, for example, \cite{semimetal_review,Gorbar:2014sja,semimetalmodel,semimetal_prediction} and references therein. The description of certain relativistic effects that may exist in Dirac and Weyl semimetals\footnote{Weyl semimetals differ from Dirac semimetals by the difference in the positions of the Fermi points for the left - handed and the right - handed fermions. In Dirac semimetals the left - handed and the right - handed Weyls fermions always come in pairs located at the same Dirac point.} was given already before the experimental discovery of Dirac semimetals \cite{semimetal_effects,semimetal_effects2,semimetal_effects3,semimetal_effects4,semimetal_effects5,semimetal_effects6,semimetal_effects7,semimetal_effects8,semimetal_effects9,semimetal_effects10,semimetal_effects11,semimetal_effects12,semimetal_effects13}. There are indications that the experimental realization of the Weyl semimetals will also be discovered  in future (see, for example, \cite{Burkov2011}).

In this paper we proceed our previous investigation of condensed matter systems with emergent relativistic invariance \cite{VolovikZubkov2014,VZ2013,Volovik:2014kja}. In the mentioned papers we concentrated on the $2+1$ D graphene. In particular, in \cite{VolovikZubkov2014,VZ2013} the emergent gravity in graphene was discussed (see also \cite{Oliva2013,Oliva2014}, where the similar results were obtained independently for the case when there are no off - plane displacements of the graphene surface). In \cite{Volovik:2014kja} the effects of emergent gravity and emergent gauge fields were investigated in the presence of dislocations (see also \cite{Guinea2014}). Here we extend our consideration to the $3+1$ D systems that simulate truly relativistic high energy theories. We concentrate on the Dirac semimetals. Those materials may be considered as elastic media \cite{Bilby1956,Kroener1960,Dzyaloshinskii1980,KleinertZaanen2004,Vozmediano2010,Zaanen2010}
and as topological matter with Weyl fermions \cite{Volovik2003,Froggatt1991,Horava2005,VZ2014NPB}. Both Cd$_3$As$_2$ and Na$_3$Bi discussed in \cite{semimetal_discovery,semimetal_discovery2,semimetal_discovery3} have two Dirac points $\pm {\bf K}$. At each Dirac point there is the pair of Weyl fermions (the right - handed and the left - handed Weyl fermions).  The chiral anomaly in Weyl and Dirac semimetals was discussed in a number of papers (see \cite{Gorbar:2013dha,chiral_torsion2,Zyuzin:2012tv,tewary} and references therein).

Since the Dirac semimetal simulates truly relativistic high - energy theories, the discussion of anomalies in the latter theories may be applied to it (with some reservations, though). In this respect it is worth mentioning that there is a certain contradiction in the results published to date. Namely, it was reported that in the presence of torsion chiral anomaly includes the topological Nieh - Yan term. In particular, this has been demonstrated using Fujikawa method \cite{chiral_torsion}. In \cite{Mielke:2006gi}, however, it was argued (this was based on the perturbative calculations with Pauli - Villars regularization) that there is no Nieh - Yan term  term in the chiral anomaly. In \cite{chiral_torsion2} the alternative derivation of the Nieh - Yan term in the anomaly was given that is based on the consideration of massless  fermions of opposite chirality as living on the boundary of the $4+1$ D topological isolator. In the present paper we do not discuss the situation, when the Nieh - Yan term may be present in chiral anomaly. Instead we restrict ourselves by the case of constant in time vielbein with vanishing components $e^i_0$ (and vanishing spin connection), in which the Nieh - Yan term vanishes.

We derive expressions for anomalies in quasiparticle currents using the explicit solution of Pauli equation in the presence of nontrivial vielbein that does not depend on time. The spectral flow in the presence of gauge fields and the torsion magnetic field results in the appearance of the anomalies in quasiparticle currents. Notice, that this method was also applied in \cite{chiral_torsion2}. However, the emergent magnetic field was not taken into account (in case, when the torsion magnetic field is present). At the same time it always accompanies the emergent gravity in Dirac semimetal. As a result the pattern of the spectral flow and the pumping of quasiparticles from vacuum that we discuss in the present paper differs somehow from that of \cite{chiral_torsion2}.

The emergent gauge and gravitational fields experienced by fermionic quasiparticles in Dirac semimetal in the presence of elastic deformations are expressed linearly in those deformations. The coefficients entering these expressions generalize the so - called Gruneisen parameter of graphene \cite{VolovikZubkov2014,VZ2013}. The elastic deformations in the presence of dislocations may be calculated using elasticity theory \cite{Landau}. It is worth mentioning, that in the presence of dislocations the emergent vielbein and emergent gauge fields are nontrivial even when the contributions of the material parameters is neglected (see also \cite{emergent_boundary}). In the present paper we assume, that the material parameters may indeed be neglected in a certain approximation, and consider the purely geometrical contributions to emergent vielbein and emergent gauge field. The spin connection in this approximation vanishes. For the early discussion of the concept of emergent gravity in the system of interacting relativistic fermions see, for example, \cite{Terazawa_gravity} and references therein.

According to our results the dislocation in Dirac semimetal as well as in graphene carries both emergent magnetic flux and  torsion singularity. The torsion singularity may be observed in the scattering of the quasipartcles on the dislocation and results in Stodolsky effect. The history of this effect originated from the suggestion of Colella and Overhauser  \cite{Overhauser} to measure the beam phase shift in the real gravitational field (that is expressed  through both gravitational constant and Plank constant). This effect was also discussed in a number of papers (see, for example, \cite{Anandan}). Stodolsky was the first who derived  \cite{stodolsky} an expression for the phase shift in the case of the gravitational field of general (non - Newtonian) form. The observation of this effect in graphene for the scattering of quasi-particle on the torsion singularity carried by the dislocation was proposed in \cite{Guinea2014}.

Finally, we calculate various effects of quantum anomalies in the quasiparticle currents in Dirac semimetals and apply our general expressions to the case, when the dislocations are present. It is worth mentioning that the anomalous production of the quasiparticles does not mean that they appear from nowhere. The microscopic currents are conserved, and the appeared charge refers to the charge of the free quasiparticles only. Suppose, that the left - handed electron at the Dirac point ${\bf K}$ appears at the dislocation (in the presence of electric field). We may consider this process as a jump of the left - handed electron from one zone (electron sea in vacuum) to another (conductance zone, the zone of quasiparticles).  The total number of electrons of each chirality remains the same, but the number of the left - handed cells in vacuum at the dislocation is decreased. Therefore, the electron that occupied a cell in vacuum before the appearance of electric field, becomes free when the electric field is turned on.

Pumping of quasiparticles from vacuum results in the appearance of chiral imbalance along the dislocation. This drives the analogue of the chiral magnetic effect that is the appearance of electric current along the dislocation. Notice, that the similar phenomenon was discussed in \cite{CME}, where the external magnetic field plays the same role as the emergent magnetic field of our case.

It is worth mentioning, that in the present paper we neglect completely Coulomb interactions between the quasiparticles. These interactions are strong due to the large effective coupling constant \cite{sarma} and break the emergent relativistic invariance. However, the investigation of graphene indicates, that such interactions may be neglected for the consideration of certain properties of the system \cite{Katsbook}. We expect, that the Aharonov - Bohm and Stodolsky effects in the scattering of the quasiparticles on the dislocations as well as the chiral anomaly driven by the emergent magnetic field incident at the dislocation are such effects that survive when the Coulomb interaction is turned on.

The paper is organized as follows. In Section \ref{sectHorava} we briefly discuss the general theory that describes  reduction of multi - fermion systems to the low energy effective models with Weyl fermions. In Section \ref{SectEmergent} we discuss the appearance of emergent gauge fields and emergent gravity with torsion experienced by fermionic quasiparticles in the Dirac semimetals in the presence of elastic deformations. In Section \ref{SectDisl} we calculate the emergent magnetic flux and the emergent torsion carried by the dislocation (assuming that the material parameters may be neglected). We also discuss how the emergent magnetic field and the torsion singularity may be observed via the Stodolsky and Aharonov - Bohm effect. In Section \ref{SectAnom} we consider quantum anomalies in quasiparticle currents in Dirac semimetal, and apply the obtained general expressions for the calculation of the anomalous production of quasiparticles at the dislocations in the presence of external electric field. Also we discuss chiral magnetic effect caused by emergent magnetic field. In Section \ref{SectConcl} we end with the conclusions.

\section{Reduction of multi - fermion systems to the models
with Weyl spinors. }
\label{sectHorava}

\subsection{Microscopic theory}

In this section we briefly discuss reduction of a multi - fermion system of general type to the model with Weyl fermions. This subject has been discussed in a number of papers (see \cite{Volovik2003} and references therein). The general theorem that describes this reduction has been proposed in \cite{Horava2005}. It may be applied to the wide class of condensed matter systems with Fermi points. The Dirac semimetal is the particular example of such a system. In this section we follow the narration of \cite{VZ2014NPB}.

 Let us denote the original
$n$ - component spinors by $\psi$. The partition function of microscopic theory has the form:
\begin{equation}
Z = \int D \psi D\bar{\psi} D\Phi {\rm exp}\Bigl(i R[\Phi] + i \int d t
\sum_{{\bf x}} \bar{\psi}_{\bf x}(t) (i\partial_t - \hat{H}(\Phi,\hat{\cal P})) \psi_{\bf
x}(t) \Bigr)\label{FIn}
\end{equation}
The collection of fields that provide interaction between the fermions is
denoted by $\Phi$. $R$ is a certain function of these fields.  In mean field approximation the
values of $\Phi$ are set to their "mean" values $\Phi_0$ and we omit integral over $\Phi$, and denote $\hat{H}(\hat{\cal P}) = \hat{H}(\Phi_0,\hat{\cal P})$.
$\sum_{{\bf x}}$ may be understood as the
integral over $d^3 x$ in continuous systems and may be considered as the sum in discrete systems.

By $\hat{\cal P}$ let us denote the three commuting Hermitian operators, whose eigenvalues parametrize the branches of the spectrum of $\hat{H}$.  Those parameters do not necessarily coincide with the ordinary three - momentum. In the basis of the corresponding eigenvectors the fermion field is denoted by $\psi_{\cal P}$.
By construction the coefficients in the
expansion of $H$ in powers of $\hat{\cal P}$ do not depend on coordinates.
Different branches of spectrum of $\hat
H$ repel each other. This means that any small perturbation pushes apart the two crossed
branches. That's why only the minimal number of branches of spectrum may
cross each other. This minimal number is fixed by momentum space topology \cite{Volovik2003}.

Let us suppose, that at ${\cal P} = {\cal P}^{(0)}$  the $n_{\rm
reduced}$ branches of $\hat H$ cross each other. Then there  exists the Hermitian matrix $\Omega({\cal P})$
such that $\tilde{H}( {\cal P}) = \Omega^+({\cal P}) \hat{H}({\cal P}) \Omega({\cal P})$ is
diagonal. Its first $n_{\rm reduced}\times n_{\rm reduced}$
block $\hat{H}_{\rm reduced}$ corresponds to the crossed branches and to the fermion components $\Psi^j_{\cal P} = \psi^j_{\cal P} , j = 1,...,n_{\rm reduced}$. All
eigenvalues of $\hat{H}_{\rm reduced}({\cal P})$ coincide at ${\cal P} = {\cal
P}^{(0)}$. The remaining block  $\hat{H}_{\rm massive}$
corresponds to the "massive" branches and to the fermion components $\Theta^j_{\cal P} = \psi^{j+n_{\rm reduced}}_{\cal P}, j = 1,...,n-n_{\rm reduced}$. The functional integral can be
represented as the product of the functional integral over "massive" modes
and the integral over $n_{\rm reduced}$ reduced fermion components.

Using transformation of the form
$\psi_{\cal P} \rightarrow e^{-i E_0 t} \psi_{\cal P}$ , $H({\cal P})
\rightarrow H({\cal P}) - E_0$ (that means an adjustment of chemical potential) we can always make the only eigenvalue of $\hat{H}_{\rm reduced}({\cal P}^{(0)})$ equal to zero.  Therefore, ${\cal P}^{(0)}$ may be considered as the position of Fermi point. At the same time variable $\cal P$ is to be treated as generalized momentum. As a result the partition
function has the form
\begin{equation}
Z = \int D \Psi D\bar{\Psi} D \Theta D\bar{\Theta} {\rm exp}\Bigl(i\int d t
\sum_{{\cal P}} \Bigl[\bar{\Psi}_{\cal P}(t) (i\partial_t - \hat{H}_{\rm
reduced}) \Psi_{\cal P}(t) + \bar{\Theta}_{\cal P}(t) (i\partial_t -
\hat{H}_{\rm massive}) \Theta_{\cal P}(t)\Bigr]\Bigr)\label{ZH2}
\end{equation}

\subsection{Reduced low energy model}

 The eigenfunctions of $\hat H_{\rm reduced}$ do not depend
on time at the Fermi point.  Therefore at low energies the integral over
$\Psi_{\cal P}$ dominates.  The other components contribute
the physical quantities with the fast oscillating factors. As a result they
may be neglected in the description of the low energy dynamics.
As it was mentioned above, the branches of spectrum repel each other. Therefore, the branches crossing occurs only if
it is protected by momentum space topology.  The minimal number of fermion components that
admits nontrivial topology is two \cite{Volovik2003} (for the discussion of topological invariants see the next subsection). Then the reduced Hamiltonian has the form
\begin{equation}
 \hat{H}_{\rm reduced} =  {\cal M}^L_k(\hat{\cal P}) \hat \sigma^k  +
{\cal M}(\hat{\cal P})  \label{Hred}
 \end{equation}
Here ${\cal M}^L_k({\cal P})$ and ${\cal M}({\cal P})$ are the real - valued functions of the generalized momentum $\cal P$. When operator $\hat{\cal P}$ is substituted instead of real - valued vector we arrive at the Hermitian operators ${\cal M}^L_k(\hat{\cal P})$ and ${\cal M}(\hat{\cal P})$.  Expression of Eq. (\ref{Hred}) is to be understood as the Hamiltonian in momentum space, i.e. in the basis of the reduced fermions (the eigenvectors of the generalized momentum are the elements of this basis). Our next purpose is to represent this Hamiltonian in the long - wave approximation in the basis of the eigenvectors of coordinates (that is we need to rewrite Eq. (\ref{Hred}) in coordinate representation). The explicit solution of this problem is not so evident in the case, when $\hat{\cal P}$ are Hermitian operators of general type. The eigenvectors of these operators that correspond to the crossed branches are expressed as complicated linear combinations of the basis vectors of coordinate representation.

According to our notations space of $\Psi_{\cal P}$ is spanned on the eigenvectors $|{\cal P},i\rangle$, $i=1,2$ (while space of  massive components is spanned on the eigenvectors $|{\cal P},i\rangle$, $i=3,...,n$), at the same time the eigenvectors of coordinates are $|{\bf x},i\rangle$, $i=1,...,n$ and we may write symbolically $\langle {\bf x}, i|\psi\rangle = \psi^i(x)$ and $\langle {\cal P}, i|\psi\rangle = \psi^i_{\cal P}$ (we should not forget, however, that $\psi$ is Grassmann variable). Our aim is to construct the linear combination of $|{\cal P},i\rangle$, $i=1,2$ in such a way, that it is the (approximate) eigenvector of the coordinate operator $\hat{x}$. We accept the natural assumption, that the definition of vectors $|{\bf x},i\rangle^{\prime}$ exists such that:

\begin{enumerate}

\item{} $\langle {\bf y}, j|{\bf x},i\rangle^{\prime} \rightarrow 0$ at $|{\bf x} - {\bf y}| \gg 1/\Lambda$ for some scale $\Lambda$;

\item{} $\langle {\bf y}, j|{\bf x},i\rangle^{\prime} \rightarrow 1$ at $|{\bf x} - {\bf y}| \ll 1/\Lambda$;

\item{} $|{\bf x},i\rangle^{\prime} = \sum_{{\cal P}; j=1,2} q_{ij}({\cal P},{\bf x}) |{\cal P},j\rangle$ for $i,j = 1,2$ with some functions $q_{ij}({\cal P},{\bf x})$.
\end{enumerate}

{\it In the other words, we may compose the wave packets of linear size $\sim 1/\Lambda$ (localized near any given point in space) out of the eigenvectors of $\cal P$ corresponding to $i = 1,2$ (that is out of the states with definite values of energy corresponding to the two crossed branches).} 
The uncertainty principle tells, that this may not be done only if the states with definite values of energy (corresponding to the two crossed branches) have distinct discrete values of at least one component of the real momentum ${\bf p}$ (its eigenvectors have the form $\sim e^{{\bf p}{\bf x}}$). We assume that our system has an infinite size and do not consider here such a marginal case.

Thus for the energies below a certain value $\Lambda$ we may define 
\begin{equation}
\Psi^i({\bf x},t) = \langle {\bf x}, i|^{\prime} \psi\rangle \equiv \sum_{{\cal P}; j=1,2} q^*_{ij}({\cal P},{\bf x}) \Psi_{\cal P}^j, \quad i = 1,2\label{vect}
\end{equation}
In this approximation we replace summation over coordinates by an integral, and write the reduced Hamiltonian as 
\begin{equation}
 \hat{H}_{\rm reduced} \approx    m^L_k(x,\hat{\bf p}) \hat \sigma^k  +
m(x,\hat{\bf p}), \label{Hred2}
 \end{equation}
where $\hat{\bf p} = - i \nabla$ is the ordinary momentum in coordinate space. Functions $m^L_k, m$ are Hermitian operators that depend both on momentum and on coordinates.
Eq. (\ref{Hred2}) should be understood as
\begin{equation}
\sum_{{\cal P}} \bar{\Psi}_{\cal P}(t)({\cal M}^L_k({\cal P}) \hat \sigma^k  +
{\cal M}({\cal P}) ) \Psi_{\cal P}(t) \approx \int d^3 {\bf x} \bar{\Psi}({\bf x},t) ( m^L_k(x,\hat{\bf p}) \hat \sigma^k  +
m(x,\hat{\bf p}))\Psi({\bf x},t)
\end{equation}
In the particular case, when  $m^L_k, m$ do not depend on coordinates the generalized momentum coincides with $\hat{\bf p} = - i \nabla$, while ${\cal M}^L_k(\hat{\cal P}) = m^L_k({\cal P})$, and ${\cal M}(\hat{\cal P})= m({\cal P})$.

\subsection{The role of momentum space topology}

The partition function is reduced to
\begin{equation}
Z = \int D \Psi D \bar{\Psi} {\rm exp}\Bigl(  i \int d^4 x
\bar{\Psi}(x) (i\partial_t - m^L_k(x,\hat{\bf p}) \hat \sigma^k  -
m(x,\hat{\bf p}))\Psi(x) \Bigr)\label{ZH__0}
\end{equation}
 We assume, that the dependence of $m^L_k, m$ on coordinates is weak and therefore in the following we may formally deal with functions  $m^L_k, m$ as independent of coordinates. This allows to introduce the notion of the {\it floating Fermi point} ${\bf K}(x)$ as the value of $\bf p$ at which $m^L_k(x,{\bf p}) \hat \sigma^k$ vanishes. The value of $\bf K$ also contains slight dependence on coordinates. When the dependence of Hamiltonian on $x$ is absent the value of $\bf K$ coincides with the position of the true Fermi point, and $m({\bf K}) = 0$ as well because in this case variable $\bf p$ coincides with $\cal P$.

The energy spectrum is $E({\cal P}) = \pm |{\cal M}^L({\cal P})| + {\cal M}(\cal P)$. If all three components of $\cal P$ are continuous, the topological invariant that protects branches crossing has the form
\begin{equation}
N= \frac{e_{ijk}}{8\pi} ~
   \int_{\sigma}    dS^i
\hat{\cal M}^L\cdot \left(\frac{\partial \hat{\cal M}^L}{\partial {\cal P}_j}
\times \frac{\partial \hat{\cal M}^L}{\partial {\cal P}_k} \right), \quad \hat{\cal M}^L =
\frac{{\cal M}^L}{|{\cal M}^L|}
\label{NH2}
\end{equation}
where $\sigma$  is the $S^2$ surface around the point ${\cal P}^{(0)}$.
For $N = 1$ the branches crossing cannot disappear as a result of the continuous deformation of the system because the zero of ${\cal M}^L_k(\hat{\cal P})$  cannot disappear while the second term ${\cal M}(\hat{\cal P})$ in Eq. (\ref{Hred}) gives the same correction to both eigenvalues of $\hat{H}_{\rm reduced}$. In this case the expansion of the functions ${\cal M}^L_k(\hat{\cal P})$  and ${\cal M}(\hat{\cal P})$ near the Fermi point has the form \cite{Volovik2003}:
\begin{eqnarray}
{\cal M}^L_i(\hat{\cal P}) &\approx& {\cal F}_i^j(\hat{\cal P}_j-{\cal P}^{(0)}_j)\,.\nonumber\\
{\cal M}(\hat{\cal P})& \approx & E_0 + {\cal F}_0^j(\hat{\cal P}_j-{\cal P}^{(0)}_j)\label{exp2}
\end{eqnarray}
Here ${\cal F}^i_j$ are the real - valued coefficients of expansion. The quantity $E_0$ may always be eliminated by the adjustment of chemical potential.  The case $N=2$ would correspond to the expansion, in which the term of the first order in $(\hat{\cal P}_j-{\cal P}^{(0)}_j)$ is absent in the expansion. The configuration of ${\cal M}^L_i({\cal P})$ with a certain  value of $N$ cannot be transformed continuously to the configuration with different value of $N$.

In the situation, when one or more components of $\cal P$ may be quantized the consideration is more involved. It is supposed, that our system of general type is obtained as a continuous deformation of a system with $m^L_k$ and $m$ independent of coordinates (in which case those functions coincide with ${\cal M}^L_k(\hat{\cal P})$  and ${\cal M}(\hat{\cal P})$) with the topological invariant of Eq. (\ref{NH2}) equal to $N=1$. In this case we are able to expand operators $m^L_i$ and $m$ in powers of $\hat {\bf p}$ near the {\it floating Fermi point}  ${\bf K}^{}_j$:
\begin{eqnarray}
m^L_i(x,\hat{\bf p}) &\approx& \frac{1}{2}\Big[f_i^j(x)(\hat{\bf p}_j-{\bf K}^{}_j(x))+(\hat{\bf p}_j-{\bf K}^{}_j(x))f_i^j(x)\Big]\,.\nonumber\\
m(x,\hat{\bf p})& \approx &K_0(x) + \frac{1}{2}\Big[f_0^j(x)(\hat{\bf p}_j-{\bf K}^{}_j(x))+(\hat{\bf p}_j-{\bf K}^{}_j(x))f_0^j(x)\Big]\label{exp}
\end{eqnarray}
Here we take into account that $m^L_i$ and $m$ are Hermitian, which results in the symmetric form of the product. By $f^j_i(x)$ and $K_0(x)$ we denote the coefficients of expansion. If their dependence on $x$ is neglected we would arrive at $f^j_i \approx {\cal F}^i_j$ and $K_0 \approx 0$.

\subsection{Emergent Weyl spinors}

Now Eq. (\ref{ZH__0}) has the form:
\begin{equation}
Z = \int D \Psi D\bar{\Psi} {\rm exp}\Bigl(i\frac{1}{2}\int d^4 x \Big[
\bar{\Psi}(x) (i\partial_t - K_0(x) - f_k^j(x)(\hat{\bf p}_j-{\bf K}^{}_j) \hat
\sigma^k )\Psi(x) + (h.c.)\Big]\Bigr)\label{ZHH}
\end{equation}
Here $(h.c.)$ means hermitian conjugated expression. According to our definition the hermitian conjugation applied to $\Psi$ gives $\bar{\Psi}$.
In  the partition function of Eq. (\ref{ZHH}),
the sum is over $k = 0,1,2,3$, and $j = 1,2,3$ while $\sigma^0 =
1$.

The coordinate dependent matrix $f^i_a({\bf x})$ coincides with the effective vielbein $e^i_a({\bf x})$ up to the factor $e({\bf x}) = {\rm det}^{1/3}\Big( f^i_a({\bf x}) \Big)$ that defines invariant integration measure over coordinates:
\begin{equation}
f^i_a({\bf x}) = e({\bf x}) \, e^i_a({\bf x})\label{fe}
\end{equation}
  We require
$e^0_a = 0$ for $a=1,2,3$, and $e \times e_0^0=1$. Here $e^{-1} = e_0^0
\times {\rm det}_{3\times 3}\, e^i_a = e_0^0$ is equal to the determinant of
the vierbein $e^i_a$.
We arrive at the partition function of the low energy effective model with emergent relativistic invariance
\begin{equation}
Z = \int D \Psi D\bar{\Psi} e^{i S[e^j_a, K_j,
\bar{\Psi},\Psi]}\label{ZI}
\end{equation}
where
\begin{eqnarray}
S &=&    \frac{1}{2} \Bigl(\int d^4 x \,e\,
\bar{\Psi}(x)  e_a^j  \hat \sigma^a   i \hat D_j  \Psi(x) +
(h.c.)\Bigr).\label{Se0}
\end{eqnarray}
Here the sum is over $a,j = 0,1,2,3$ while $\sigma^0 \equiv 1$, and $\hat
D$ is the covariant derivative that includes the $U(1)$ gauge field $K_\mu = (K_0,{\bf K})$.

If we want to take into account interaction between the fermions due to the fluctuations of the fields $\Phi$ in Eq. (\ref{FIn}), then we should start from Eq. (\ref{FIn}) with the fixed value of $\Phi$ that is not equal to its mean value $\Phi_0$. At each value of $\Phi$ we proceed through the above steps and come to Eqs. (\ref{ZI}), (\ref{Se0}) with the values of $e^j_a$ and $K_\mu$ dependent on $\Phi$. This dependence may be rather complicated. Nevertheless, in principle, the expansion in powers of $\Phi - \Phi_0$ with phenomenological coefficients may be used. This will bring us to the theory of relativistic fermions interacting with the collection of bosonic fields.

One can see, that in the considered long wave approximation the emergent
spin connection $C_{\mu}$ does not arise.  Actually, we may add the spin connection for completeness. But for the Hermitian Hamiltonian it will effectively result in the interaction of the Weyl fermion with a new $U(1)$ gauge field and to the correction to chemical potential. Thus, those contributions may be absorbed by the field $K_\mu$ and by the overall chemical potential.

\section{Emergent Weyl fermions in Dirac semimetal in the presence of elastic deformation. }
\label{SectEmergent}
\subsection{Dirac semimetals as elastic media}

First of all let us consider the Dirac semimetal as elastic media.  The deformations in $3D$ elastic media are described by the change in the coordinates of atoms $X^i({\bf x})=u^i({\bf x}) + x^i$, where $u^i({\bf x})$ is the displacement field, which describes the displacement of atoms from their equilibrium positions. We assume that the elastic deformations are small
 $\partial_i u^k\ll 1$.

The displacement field is expressed in terms of the elastic deformations and rotations:
\begin{eqnarray}
&&\partial_j u_i(x) = \epsilon_{ij}(x) + \omega_{ij}(x)
 \label{elasticity3}
\\
&& \epsilon_{ij}= \frac{1}{2}(\partial_ju_i + \partial_i u_j ) , \quad    \omega_{ij}= \frac{1}{2}(\partial_ju_i - \partial_i u_j )
 \label{elasticity4}
\end{eqnarray}
Tensor $\omega_{ij}$ describes local rotations, while tensor of elastic deformations is denoted by $\epsilon_{ij}$.
Throughout the text we assume that elastic deformations do not depend on time.

Elasticity theory operates with metric
 \begin{equation}
g_{ik}= \frac{\partial X^l}{\partial x^i}
\frac{\partial X^l}{\partial x^k} \approx \delta_{ik} + 2\epsilon_{ik}
 \label{elasticity}
\end{equation}
The naive supposition is that the fermions propagate in Riemannian space corresponding to this metric. However, we will see below, that as well as in graphene \cite{VolovikZubkov2014,VZ2013}, the fermionic quasiparticles in Dirac semimetal in the presence of elastic deformations propagate in Riemann - Cartan space with vanishing curvature (Weitzenbock space), which is determined by a certain vierbein. In leading approximation this vierbein is expressed linearly through $\epsilon_{ij}$.

\subsection{Four Weyl fermions in Dirac semimetal.}

It follows from the microscopic theory \cite{semimetal_review,Gorbar:2014sja,semimetalmodel,semimetal_prediction} that in the considered Dirac semimetals (without strain) there are four Weyl fermions \cite{semimetal_discovery,semimetal_discovery2,semimetal_discovery3} (the pair left - handed fermion -- right - handed fermion at ${\bf K}^{(0)}$ and the pair left - handed fermion -- right - handed fermion at $-{\bf K}^{(0)}$). Thus in the absence of elastic deformations we have two pairs of coinciding Weyl points $\pm{\bf K}^{(0)}$ and the low energy effective action of the form
\begin{equation}
S =\frac{1}{2} \sum_{s= L,R}\, \sum_{\pm} \int d^3x dt \, e^{(0)}\,\bar{\Psi}_{s,\pm} e_K^{(0)\mu}\sigma_s^K \Big(i \partial_{\mu} + K^{(0)\pm}_{\mu}\Big)\Psi_{s,\pm} + (h.c.),\quad \sigma^0=1\label{effact0}
\end{equation}
where $\sigma^K_R = \sigma^K$ while $\sigma^K_L = \bar{\sigma}^K$, and $\bar{\sigma}^a = - \sigma^a$ for $a = 1,2,3$, while $\bar{\sigma}^0 = \sigma^0 = 1$. $\Psi, \bar{\Psi}$ are the independent fermion Grassmann variables that describe quasi - particles living near the  Weyl points $\pm{\bf K}^{(0)}(x)$.
Here ${ K}^{(0)\pm}_k = \pm {\bf K}_k^{(0)}$ for $k = 1,2,3$ and  ${ K}^{(0)\pm}_0 = 0$ while
\begin{equation}
e^{(0)}=v_F, \quad e^{(0)i}_a =  \hat{f}^i_a,\quad e^{(0)i}_0 =0, \quad e^{(0)0}_a =\frac{1}{v_F} \delta^0_0, \quad a,i,j,k=1,2,3
 \label{Connection00}
\end{equation}
where $v_F \hat{f}^i_a$ has the meaning of anisotropic Fermi velocity.
{\it The considered Dirac semimetals \cite{semimetal_discovery,semimetal_discovery2,semimetal_discovery3}  are anisotropic, that results in the anisotropic Fermi velocity corresponding to $3\times 3$ matrix $\hat{f} = {\rm diag}(\nu^{-1/3},\nu^{-1/3},\nu^{2/3})$ with $\nu \ne 1$. }

We do not exclude, that in the presence of elastic deformations the positions of the two Dirac points may split into four (floating) Weyl points depending on the position in space. We denote them by  ${\bf K}^{L,\pm}$ and ${\bf K}^{R,\pm}$. The definition
of ${\bf K}$ depends on the choice of coordinate system, i.e. on the parametrization of the semimetal.  We may choose the parametrization, in which the coordinates of atoms in the crystal are the same as in the unperturbed lattice. In this reference frame we have ${\bf K}^{L,\pm}   \approx   \pm {\bf K}^{(0)} + {\bf A}^{(a)L,\pm}$ and ${\bf K}^{R,\pm}   \approx   \pm {\bf K}^{(0)} + {\bf A}^{(a)R,\pm}$, where ${\bf A}^{(a)L,\pm}$ and  ${\bf A}^{(a)R,\pm}$ are to be interpreted as the emergent $U(1)$ gauge fields. We refer to the given parametrization as to the {\it accompanying reference frame}.

If the crystal is deformed only slightly, the other parametrization is preferred, which is called typically the {\it laboratory reference frame}. In this parametrization the coordinates of atoms are their real $3D$ coordinates. Then the transformation between the two parametrizations is given by:
\begin{equation}
X^k({\bf x})=u^k({\bf x}) + x^k, \quad k = 1,2,3
\end{equation}
Here the laboratory reference frame coordinates are denoted by $X^k$ while the coordinates of the accompanying  reference frame are denoted by $x^k$.

We neglect possible interactions between the fermions located at different Weyl points. Therefore, the general theory described in the previous section may be applied to each of the mentioned Weyl fermions. As a result the effective low energy action has the form of the sum over the four Weyl points:
\begin{equation}
S =\frac{1}{2} \sum_{s= L,R}\, \sum_{\pm} \int d^3x dt \, e\,\bar{\Psi}_{s,\pm} e_K^{s,\pm,\mu}\sigma_s^K \Big(i \partial_{\mu} + K^{s,\pm}_{\mu}\Big)\Psi_{s,\pm} + (h.c.),\quad \sigma^0=1\label{effact}
\end{equation}
$\Psi, \bar{\Psi}$ are the independent fermion Grassmann variables that describe quasi - particles living near the (floating) Weyl points ${\bf K}^{s,\pm}(x)$.
 The effective $3+1$ D  vielbein $e^{s,\pm,\mu}_a$ may be different for the four Weyl fermions. Therefore, we add the superscript $^{s,\pm}$. Here ${K}^{s,\pm}_k =  {\bf K}_k^{s,\pm}$ for $k = 1,2,3$, while ${ K}^{s,\pm}_0$ may be nonzero.

\subsection{Expressions for emergent vielbein and emergent gauge field.}
\label{Secthom}

For small elastic deformations the values of emergent gauge field and the values of the vielbein should be expressed through the derivatives of the displacement vector $u^k$. In the decomposition of Eq. (\ref{elasticity3}) the contribution of $\omega_{ij}$ corresponds to rotations that cannot cause strain and therefore does not contribute to the emergent vielbein and to emergent gauge field in accompanying reference frame. In laboratory reference frame $\omega_{ij}$ may appear, however. The explanation of this difference follows the analogy with graphene \cite{VolovikZubkov2014,VZ2013}, where it results from the consideration of microscopic theory.
Explicitly, in Dirac semimetals as well as in graphene the microscopic model operates with the fermion operators $\psi(x)$ defined on the sites of the lattice (see \cite{semimetal_review,Gorbar:2014sja,semimetalmodel,semimetal_prediction} and references therein). These operators create and annihilate electrons on certain orbitals that belong to the atoms located at the lattice sites. The microscopic Hamiltonian depends on real distances between the atoms through its phenomenological parameters. Small elastic deformations result in small variations of these parameters.

The accompanying reference frame is distinguished because in this reference frame the coordinates of atoms are the same as in the unperturbed lattice. This is this reference frame, which appears naturally in the low energy effective continuous model because operators $\psi(x)$ are attached to the sites of the lattice. Therefore, this is this reference frame, in which the parameters of effective action Eq. (\ref{effact}) $e^\nu_a$ and $K_\mu$ are equal to their unperturbed values plus corrections that are proportional to the derivative of the displacement vector $u^i$. As it was mentioned above, tensor $\omega_{ij}$ in the decomposition of Eq. (\ref{elasticity3}) corresponds to rotations. Rotations of crystal cannot change the distances between the atoms. Therefore, $\omega_{ij}$ does not enter the expressions for $e^\nu_a$ and $K_\mu$ in this reference frame. For the mentioned above reasons this consideration refers to the accompanying reference frame only and does not work, for example, in laboratory reference frame, in which the coordinates of atoms are given by their real $3D$ coordinates. In laboratory reference frame the rotation of crystal results in rotation of Fermi point ${\bf K}$ (while in accompanying reference frame rotations cannot change $\bf K$). In the same way in laboratory reference frame rotations of crystal result in rotation of vector $e^i_a$ for each $a = 0,1,2,3$. It will be seen further, that in Laboratory reference frame tensor $\omega_{ij}$ enters expressions for the emergent vielbein and emergent gauge field due to the transformation between the two reference frames, which contains both $\omega_{ij}$ and $\epsilon_{ij}$.

Thus, the most general leading linear dependence of gauge field on $\epsilon_{ij}$ in  accompanying reference frame has the form
\begin{eqnarray}
{\bf A}^{(a)L,R,\pm}_i&=& \frac{1}{a}\beta^{L,R,\pm}_{ijk} \epsilon_{jk},\quad A_0^{(a)L,R,\pm}= \frac{1}{a}\beta^{L,R,\pm}_{0jk} \epsilon_{jk}, \quad i,j,k=1,2,3
 \label{TetradGaugeEffectiveA0}
\end{eqnarray}
where ${\bf K}^{L,R,\pm}_k = \pm {\bf K}_k^{(0)}+  {\bf A}^{(a)L,R,\pm}_k$ for $k = 1,2,3$ and  ${ K}^{L,R,\pm}_0 = {\bf A}^{(a)L,R,\pm}_0$, while $\beta^{L,R,\pm}_{ijk}$ is the tensor of dimensionless material parameters (similar to the Gruneisen parameter of graphene), $a$ is the lattice spacing.

In laboratory reference frame the values of ${\bf K}$  are given by:
\begin{eqnarray}
{\bf K}^{\pm}_i & \approx & \frac{\partial x^k}{\partial X^i} \Big(\pm {\bf K}_k^{(0)}+  {\bf A}^{(a)\pm}_k\Big) \approx \Big(\delta_i^k -  \partial_i u^k \Big) \Big(\pm {\bf K}_k^{(0)}+  {\bf A}^{(a)\pm}_k\Big)\nonumber\\ & \approx &  (\pm {\bf K}_i^{(0)}\mp \nabla_i({\bf u}\cdot {\bf K}^{(0)})) +    {\bf A}^{(a)\pm}_i
 \label{DiracPosition1}
\end{eqnarray}
Here we omit the superscripts $L,R$. The field ${\bf A}^{(a)}$ is given by Eq. (\ref{TetradGaugeEffectiveA0}). In the following we denote the emergent gauge field in laboratory reference frame by
\begin{eqnarray}
 {\bf A}^{\pm}_i & \approx &  \mp  \nabla_i({\bf u}\cdot {\bf K}^{(0)})) +    {\bf A}^{(a)\pm}_i
 \label{DiracPosition1_}
\end{eqnarray}
The first term in the right hand side of this expression contains the geometric contribution, which comes from the coordinate transformation of the original position ${\bf K}^{(0)}$ of the Dirac point in the non-deformed lattice.

In the same way we relate parameters $f^i_a = e\, e^i_a $ of the fermionic spectrum with the elastic deformation.
As well as for ${\bf K}$ the definition of $f^i_k$ depends on the choice of coordinate system, i.e. on the parametrization of the semimetal. In the accompanying reference frame we have:
\begin{equation}
 f^i_a \approx v_F \left(\hat{f}^i_a-\gamma^i_{a jk} \epsilon_{jk} \right),\quad f^i_0 \approx -v_F \gamma^i_{0 jk} \epsilon_{jk}, \quad f^0_a = \delta^0_0, \quad a,i,j,k=1,2,3
 \label{Connection0}
\end{equation}
where at each Weyl point the tensor of new dimensionless material parameters $\gamma^i_{ajk}$ is defined (we omit here the superscripts $L,R,\pm$).


Applying coordinate transformation to Eq. (\ref{Connection0}) we come to the following expression for $f^i_a$ in the laboratory reference frame:
\begin{equation}
 f^i_a ={\rm det}^{-1} \Big(\frac{\partial X}{\partial x}\Big) v_F\left(\hat{f}^k_a-\gamma^k_{a jl} \epsilon_{jl} \right)\,\frac{\partial X^i}{\partial x^k}\approx v_F\left(\hat{f}^i_a(1-\partial_k u_{k}) + \hat{f}^k_a\partial_k u^i - \gamma^k_{a jl} \epsilon_{jl} \right)
 \label{Connection1}
\end{equation}

Then we come to
\begin{equation}
 f^i_a \approx v_F\left(\hat{f}^i_a(1-\epsilon_{kk}) +  \hat{f}^k_a\epsilon_{ia} + \hat{f}^k_a\omega_{ia}-\gamma^i_{a jl} \epsilon_{jl} \right),  \quad    \omega_{ij}= \frac{1}{2}(\partial_ju_i - \partial_i u_j )
 \label{Connection11}
\end{equation}
The terms $\hat{f}^i_a(1-\partial_k u_{k}) + \hat{f}^k_a\partial_k u^i$ in the right hand side of Eq.(\ref{Connection1}) are of the geometric origin: they come from the coordinate transformation from the original non-disturbed to the strained semimetal.

We come to the conclusion, that when the material parameters $\beta_{ijk}$ and $\gamma^i_{a jk}$ are neglected, in laboratory reference frame we arrive at the same expression for the vielbein for all four Weyl fermions that is given by the pure geometric contribution. In this approximation (which will be assumed in the following) the spin connection vanishes while the emergent gauge field is again the same for all four Weyl fermions, is given by the first term of Eq. (\ref{DiracPosition1_}) and has opposite charges for opposite Dirac points $\pm {\bf K}^{(0)}$.

The contributions to various physical quantities of the material parameters may be neglected if those dimensionless parameters are small. {\it In the following we assume that for the considered Dirac semimetals the values of $\beta_{ijk}$ and $\gamma^i_{a jk}$ are indeed small.} This means that elastic deformations with $\epsilon_{ij} \sim 1$ result in the contributions to the quasiparticle Hamiltonian that are small compared to the unperturbed Hamiltonian. More explicitly, first, the elastic deformations should result in the contribution to the emergent gauge field that is small compared to its unperturbed value ${\bf K}^{(0)}$. This is a very natural assumption because at the same time it is required in order to apply low energy effective field theory. (It is supposed, that this theory may be applied if values of momenta including its shift due to the emergent gauge field are small compared to ${\bf K}^{(0)}$.) Second, elastic deformations should give contributions to the anisotropic Fermi velocity that are small compared to its unperturbed values. Qualitatively this is also natural, because this means that elastic deformations are not able to change considerably speed of various excitations existing in the semimetal. Nevertheless, the mentioned assumption is to be checked experimentally for the particular materials.

\section{Emergent geometry in the presence of dislocations}
\label{SectDisl}


\subsection{Accompanying reference frame}

Effective geometry experienced by fermions acquire singular contributions to torsion concentrated at the origin $l^0$ of the dislocation (that is the closed line or a line connecting the boundaries). At the origin $l^0$ of the dislocation the cells of the crystal may be substituted by irregular cells. Then the extra sequence of regular cells is added along the surface  $\cal J$ (line $l^0$ is its boundary). For the low energy effective theory this results in cutting of the volume along the surface $\cal J$ that begins at $l^0$ and goes to infinity. Then the strip of a finite width is added along $\cal J$.  The resulting volume is sewn along the cut. As a result in the accompanying reference frame (where the coordinates of the atoms are the same as in the unperturbed lattice) there is the uncertainty in the definition of the parametrization at the cut. This uncertainty gives nonzero value to the following integral along the contour $\cal C$ surrounding $l^0$:
\begin{equation}
b^i = \int_{\cal C} d x^i
\end{equation}

\subsection{Laboratory reference frame}

In laboratory reference frame the parametrization of semimetal volume does not contain ambiguity. The ambiguity in the parametrization of accompanying reference frame appears through the displacement field defined as a function of $X^k$:
\begin{equation}
x^k = X^k - u^k(X)
\end{equation}
In accompanying reference frame the parametrization is defined modulo the step - like discontinuity at the cut. Therefore, although the displacement field $u^i(X)$ has a step - like discontinuity concentrated along the cut, its derivative $\partial_j u^k$ is continuous. The Burgers vector is
\begin{equation}
b^i = \int_{\cal C} d x^i = - \int_{\cal C} d u^i
\end{equation}
For $l^0$ of the form of the infinite straight line we may choose, for example, the following representation:
\begin{equation}
u^a = -  \phi \frac{b^a}{2\pi}  + u^a_{\rm cont},\label{using}
\end{equation}
where $\phi$ is the polar angle in the plane orthogonal to the dislocation\footnote{if the third axis is directed along the dislocation, then $X_1= X^0_1 + r \, {\rm cos} \, \phi$, $X_2=X^0_2+ r \, {\rm sin} \, \phi$} while $u^a_{\rm cont }$ is continuous along the cut (i.e. it does not contain the discontinuity along the cut). However, $u_{\rm cont}$ may be undefined at the origin of the dislocation.
In the following we will use the laboratory reference frame, where there is no ambiguity in parametrization.

\subsection{Elasticity equations}
\label{sectelast}

 Elastic part of the free energy may be written as \cite{Landau}:
\begin{equation}
F = \frac{1}{2}\int d^2 X \Big(\lambda \epsilon^2_{ll} + 2 \mu \epsilon_{ik} \epsilon_{ik} \Big), \label{F}
\end{equation}
where $\lambda$ and $\mu$ are Lame coefficients.
We may neglect the term quadratic in $u_a$ in $\epsilon_{ij}$. Therefore, the free energy is quadratic in $u_a$ ($a=1,2$). As a result the differential equations are linear in $u^a$.

The variation over $u^a$ ($a=1,2$) gives:
\begin{equation}
\partial_k \epsilon_{ik} + \frac{\sigma}{1-2\sigma} \partial_i \epsilon_{ll} = 0,\label{elaste}
\end{equation}
\revision{Here we introduced  Poisson parameter $\sigma$ as:
$\frac{\sigma}{1-2\sigma}=\frac{\lambda}{\mu}$.}   Eq. (\ref{elaste}) together with the boundary conditions determines the values of the equilibrium displacement field $u^a$. Let us consider the case, when $l^0$ is infinite straight line. The problem becomes two - dimensional. Its solution is given in \cite{Landau}. For the completeness we present it in Appendix A.
The resulting value of $u^k_{\rm cont}(X)$ is given by
\begin{eqnarray}
 u^k_{\rm cont}(X) &=&  -\frac{1-2\sigma}{2(1-\sigma)}  b^l {\bf i}^{kl}   \frac{1}{2\pi}  {\rm log} \frac{|X|e^{\gamma}}{2 R}+  b^l {\bf i}^{il} \frac{1}{1-\sigma} \frac{\hat{X}^i \hat{X}^k}{4\pi} \label{ucontsol1}
\end{eqnarray}
This expression should be regularized both at $|X|\rightarrow 0$ and at $|X| \rightarrow \infty$. The considered effective field theory fails at the distances of the order of the lattice spacing $a$. The effects of the finite size of the sample are also strong. It is worth mentioning, that while the values of $u_{\rm cont}^k$ may be large, its derivatives are small for sufficiently small $b$ because after the differentiation the expression in Eq. (\ref{ucontsol1}) tends to zero at $|X|\rightarrow \infty$.

\subsection{Emergent magnetic field carried by the dislocation}

In order to calculate the emergent magnetic field we use integral expression
\begin{equation}
{\bf i}^{ijk}\int_{\cal S} H^i dx^j \wedge dx^k \equiv \int_{\partial {\cal S}} A_k dX^k \label{AdefI}
\end{equation}
For the considered solution of elasticity equations   we represent the right hand side of this expression as follows
\begin{equation}
\int_{\partial {\cal S}} A_k dX^k =  \pm b^i {\bf K}^{(0)}_i + O(\beta)\label{AdefI2}
\end{equation}
The first term in this expression gives the following singular contribution to magnetic field:
\begin{equation}
H^k_{{\rm sing}} \approx \pm b^i {\bf K}^{(0)}_i \int_{l^0}  d Y^k(s) \delta^{(3)} (X-Y(s)),\label{Hsing}
\end{equation}
The unperturbed Fermi point is defined up to the transformation ${\bf K}^{(0)} \rightarrow {\bf K}^{(0)} + {\bf G}$, where ${\bf G}$ is the vector of inverse lattice. This corresponds to the change of the magnetic flux by
$\Delta \Phi={\bf b}\cdot{\bf G} = 2\pi N$, where $N$ is integer. Such a change of the magnetic flux is unobservable for Weyl fermions.

According to the remark presented at the end of Section \ref{SectEmergent}, we assume, that the mentioned singular contribution to magnetic flux dominates over the regular one proportional to parameters $\beta_{ijk}$. That means that the dimensionless generalized Gruneisen parameters $\beta_{ijk}$ are small for all four Weyl points. This supposition is also justified by the analogy with graphene. We also suppose, that the dimensionless parameters $\gamma^i_{ajk}$ are small, and the corresponding contributions to torsion may be neglected compared to the singular one considered in the next subsection.

\subsection{Torsion carried by the dislocation}

In the absence of dislocation the torsion tensor is defined as
\begin{eqnarray}
T^a_{jk} &\equiv & \partial_{[j}  e^a_{k]},
\end{eqnarray}
where $e^a_k$ is the (dimensionless) inverse $3+1$ D vielbein related to $f_a^i$ according to Eq. (\ref{fe}).
In the presence of dislocation we use integral representation:
\begin{equation}
\frac{1}{2}\int_{\cal S} T_{ij}^a dX^i \wedge dX^j \equiv \int_{\partial {\cal S}} e^a_k(X) dX^k \label{TedefI}
\end{equation}
In order to calculate torsion at $X^0$ we should choose $\cal S$ as its small vicinity.

In the accompanying reference frame at each Weyl point for the components of $e^a_k$ with $k,a = 1,2,3$ we have:
\begin{eqnarray}
e^a_k(x) &=& \Big(\hat{f}^k_a -\gamma^k_{a jl} \epsilon_{jl}\Big)^{-1}{\rm det}^{1/3}\Big(\hat{f}^k_a -\gamma^i_{a jk} \epsilon_{jk}\Big)\nonumber\\ &\approx & \hat{r}^a_k + O(\gamma), \quad \hat{r}^a_k = \Big[\hat{f}^{-1}\Big]^a_k = {\rm diag}(\nu^{1/3},\nu^{1/3},\nu^{-2/3})
\end{eqnarray}
The values of $e^a_k$ in laboratory reference frame are given by
\begin{eqnarray}
e^a_k(X) = \frac{\partial x^j}{\partial X^k} e^a_j(x) &\approx & \Big(\hat{r}^a_k -  \hat{r}^a_j\partial_k u^j  + O(\gamma)\Big)\label{elab}
\end{eqnarray}
In principle, we should also transform derivatives entering $ \epsilon_{ka}$: $\frac{\partial}{\partial x^k} \rightarrow \frac{\partial X^i}{\partial x^k}\frac{\partial}{\partial X^i}$. However, we are able to substitute here $\frac{\partial X^i}{\partial x^k}$ by $\delta^i_k$ in the approximation linear in displacement $u^i, i = 1,2,3$.

According to Eq. (\ref{TedefI}) torsion is related to the circulation of the inverse vielbein $e^a_k$ along the closed contour.   For the case, when $\partial_k u^a \partial_l u^a$ ($a=1,2,3$) may be neglected we have:
\begin{eqnarray}
\int e^a_k(X)d X^k &\approx & \int \Big( - \hat{r}^a_j\partial_ku^j(X) + O(\gamma)\Big) dX^k = \hat{r}^a_j b^j + O(\gamma) \approx \tilde{b}^a \label{eq}
\end{eqnarray}
Recall that $\hat{r}^a_k  = {\rm diag}(\nu^{1/3}, \nu^{1/3}, \nu^{-2/3})$ and $\hat{f}^k_a  = {\rm diag}(\nu^{-1/3}, \nu^{-1/3}, \nu^{2/3})$.  Then $\tilde{b}^a = \hat{r}^a_k b^k$. We arrive at the following singular contribution to torsion:
\begin{eqnarray}
T^a_{ij, {\rm sing}} &\approx & \tilde{b}^a {\bf i}_{ijk}\int_{l^0}  d Y^k(s) \delta^{(3)} (X-Y(s)) + O(\gamma)\label{tsing}
\end{eqnarray}

\subsection{Observation of torsion singularity and emergent magnetic flux}

Here and below we use the following equivalent notations for the four vectors: $X^{\mu}=(X^0,X^1,X^2,X^3)=(t,{\bf r})$.
Let us estimate the effect of torsion singularity on the scattering of quasiparticles on the dislocation.  The wave function satisfies Pauli equation
\begin{equation}
\frac{1}{2} \sigma_{L,R}^K \Big( e \, e_K^{\mu} i D_{\mu} +  i D_{\mu} e\, e_K^{\mu}\Big)\Psi = 0,  \quad \mu,K = 0,1,2,3\quad \sigma^0 = 1 \label{pauli}
\end{equation}
The covariant derivative $D_{\mu}$ contains both emergent magnetic field and emergent spin connection.
Recall, that the vielbein is given by:
\begin{eqnarray}
 e_a^{\mu} & = & \hat{f}_a^{\mu} + \hat{f}^\nu_a \partial_\nu u^{\mu} + O(\gamma) , \quad a,\mu = 1,2,3 \quad e_0^{\mu}  = \frac{1}{v_F}(1+O(\gamma))\delta_0^\mu\nonumber\\
 e^a_{\mu} & = & \hat{r}^a_{\mu} - \hat{r}^a_\nu \partial_\mu u^{\nu} + O(\gamma), \quad a,\mu = 1,2,3  \quad e_\mu^{0}  = {v_F}(1+O(\gamma))\delta^0_{\mu}\nonumber\\
 e & = & {\rm det} \, e^a_{\mu} = v_F(1 - \partial_i u^i + O(\gamma)), \quad i = 1,2,3
\end{eqnarray}
 In the following we neglect all terms linear in material parameters $\gamma,\beta$.
 Let us represent
\begin{equation}
\Psi({\bf r},t)=\Psi_{\bf k}({\bf r},t) e^{-i \int^{(t,{\bf r})}_{(t_0,{\bf r}_0)} k_I e^I_{\mu}(Y) d Y^\mu + i \int^{(t,{\bf r})}_{(t_0,{\bf r}_0)} A_{\mu}(Y) d Y^\mu}
\,
 \label{psi}
\end{equation}
where $k_\mu=(k_0,k_x,k_y,k_z)$. We omit the superscripts $\pm$ that mark the valleys for simplicity. The integral here is along the trajectory ${\cal C}^{({\bf r},t)}_{({\bf r}_0,t_0)}$ given by the function $Y^{\mu}(s,X)$. It is parametrized by $s$ and depends on the endpoint $X^{\mu} = (t,{\bf r})$. We assume, that $\Psi_k$ is slow varying. {Then vector $k_a$ plays the role of the three - momentum of the incoming particle because around the point $(t_0,{\bf r}_0)$ situated far from the dislocation the vielbein $e^a_{\mu} \approx {\rm diag}(v_F, \nu^{1/3},\nu^{1/3},\nu^{-2/3}) $ is flat, and
\begin{equation}
\Psi \sim e^{-i v_F k_0 (t-t_0)  + i  {\bf p}_j ({\bf r}^{j}-{\bf r}^{j}_0)}
\end{equation}}
Here the eigenvalue of the $3$ - momentum $-i \nabla$ is denoted by ${\bf p}_j = - \hat{r}^a_j k_a$.  Eq. (\ref{psi}) actually defines a series of different solutions that correspond to the difference in a winding of ${\cal C}$ around the dislocation. Phase $i \int^{({\bf r},t)}_{({\bf r}_0,t_0)} k_a e^a_{\mu}(Y) d Y^\mu$ in Eq. (\ref{psi}) may be interpreted as follows: the translation between the two points $X_0=(t_0,{\bf r}_0)$ and $X=(t,{\bf r})$ in Weitzenbock space is defined as  $R^a=\int^{(t,{\bf r})}_{(t_0,{\bf r}_0)} e^a_{\mu}(Y) d Y^\mu$. Plane wave is determined by phase $k_a R^a$. Substituting Eq. (\ref{psi}) to Eq.(\ref{pauli}) one obtains
\begin{equation}
\Big(\sigma_{L,R}^I k_I +  \sigma_{L,R}^I    e_I^{\mu}\Big[ i \partial_{\mu}  + \int^{(t,{\bf r})}_{(t_0,{\bf r}_0)} \Big(k_a T^a_{\nu\rho}(Y) - F_{\nu\rho}(Y)\Big)\frac{\partial Y^{\nu}(s,X)}{\partial X^{\mu}} d Y^\rho  + i \Gamma_\mu \Big]\Big) \Psi_{\bf k}({\bf r},t)= 0 \,,
\end{equation}
Here we denote $\Gamma_\mu = \frac{1}{2 \, e} e^a_\mu \nabla_{\nu} e\, e_a^{\nu} = \frac{1}{2} e_a^{\nu}T^a_{\mu\nu}$.

 We can always choose the trajectories $Y(s,X)$ far from the origin of the dislocation (at the distance $|{\bf r}| \gg a$.) For the vielbein given by Eq. (\ref{elab}) with $u^a$ given by Eq. (\ref{ucontsol1}) we have  $\nabla_{\mu} e\, e_a^{\mu} \sim \frac{1}{|{\bf r}|^2}$. The non - singular part of torsion is also $\sim \frac{1}{|{\bf r}|^2}$.  Therefore, $\Psi_k$ is constant and satisfies
  $\sigma^I_{L,R} k_I \Psi_k = 0$.

The infinite straight line $l^0$ of the dislocation plays the role of a line - like hole in $3D$ volume of the semimetal (or a 2D surface in $3+1$ D space - time) carrying singularity of torsion $T_{12}^a = \tilde{b}^a + O(\gamma)  \delta^{(2)}(X)$. The difference between the phases of the two solutions defined by paths ${\cal C}^{(i)}$ and ${\cal C}^{(j)}$ ended at the same point $X^{\mu}$ is defined by the winding number $K_{ij}$ of the contour ${\cal C}^{(ij)} = {\cal C}^{(i)} - {\cal C}^{(j)}$ around $l^0$. The two solutions differ by the phase factor:
 \begin{equation}
\Psi^{(i)}_k(X) = \Psi^{(j)}_k e^{-i \int_{{\cal C}^{(ij)}} e_{\mu}^a k_a dX^{\mu} +  i \int_{{\cal C}^{(ij)}} A_{\mu} dX^{\mu} } =  \Psi^{(j)}_k e^{ i K^{ij} b^k {\cal K}_k  + O(\gamma,\beta,\lambda)  }, \quad {\cal K}_k = \pm{\bf K}^{(0)}_k  + {\bf p}_k \label{psi2}
\end{equation}
Here ${\cal K}$ may treated as the total momentum of the quasiparticle (that is the sum of the unperturbed Dirac point $ \pm {\bf K}^{(0)}$ and ${\bf p}$). Following \cite{Volovik:2014kja} we call the appearance of this phase in the wave function (proportional to the the winding number and to the momentum of particle) the Stodolsky effect. The observation of this effect in graphene was first proposed in \cite{Guinea2014}. Actually, our consideration of the present section follows the approach of \cite{Guinea2014} in its form developed in \cite{Volovik:2014kja}.

\section{Effects of anomaly in the presence of dislocations}

\label{SectAnom}

\subsection{General expression for the anomaly}

The partition function of the effective low energy field theory of Dirac semimetal has the form:
\begin{equation}
Z= \int D\bar{\Psi}D\Psi\, {\rm exp}\Big(i\frac{1}{2} \sum_{s=L,R}\sum_{\pm} \, \int d^4 X \,\Bigl[ e^{s,\pm}\,\bar{\Psi}_{s,\pm} e_K^{s,\pm,\mu}\sigma_s^K \Big(i \partial_{\mu} + K^{s,\pm}_{\mu}-B_\mu\Big)\Psi_{s,\pm} + (h.c.)\Bigr]\Big)
\end{equation}
We introduced here the external $U(1)$ gauge field $B$, take into account that $K^{s,\pm}_{\mu} =  \pm K^{(0)}_{\mu} + A^{s,\pm}_{\mu}$ and assume that $e_a^{\mu} \approx {\rm diag}(1/v_F, \nu^{-1/3},\nu^{-1/3},\nu^{2/3})$.

We perform the following coordinate transformation:
\begin{eqnarray}
X^0\rightarrow v_F^{-1} X^0, \quad X^{1,2} \rightarrow \nu^{-1/3} X^{1,2},\quad  X^3 \rightarrow \nu^{2/3} X^3
\end{eqnarray}
and the corresponding transformation of the other vectors and tensors (with the space - time indices $\mu,\nu,...$) including $K_\mu$ and $u^\mu$.
Let us neglect all effects that result from material parameters $\beta,\gamma$. Then, in the presence of the dislocation we get:
\begin{eqnarray}
e_a^{\mu} & = & \delta_a^{\mu} + \partial_a u^{\mu} + O(\gamma) , \quad a,\mu = 1,2,3 \quad e_0^{\mu}  = (1+O(\gamma))\delta_0^\mu\nonumber\\
 e^a_{\mu} & = & \delta^a_{\mu} - \partial_\mu u^{a} + O(\gamma), \quad a,\mu = 1,2,3  \quad e_\mu^{0}  = (1+O(\gamma))\delta^0_{\mu}\nonumber\\
 e & = & |{\rm det} \, e^a_{\mu} | = (1 - \partial_i u^i+ O(\gamma)), \quad i = 1,2,3\nonumber\\
 A^{\pm}_{i} &=&  \mp \nabla_i({\bf u}\cdot {\bf K}^{(0)}) + O(\beta), \quad A_0 = O(\beta)
\end{eqnarray}
This form of the emergent fields allows us to rewrite the partition function as follows:
\begin{equation}
Z \approx \int D\bar{\Psi}D\Psi \, {\rm exp}\Big( i\frac{1}{2} \sum_{s=L,R}\sum_{\pm} \, \int d^3 X dt \, e\,\Big[\bar{\Psi}_{s,\pm} e_K^{\mu}\sigma_{s}^K \Big(i \partial_{\mu} \pm K^{(0)}_{\mu}+ A^{\pm}_\mu - B_\mu \Big)\Psi_{s,\pm} + (h.c.)\Big]\Big),
\end{equation}

Let us introduce the left - and the right - handed currents $J_J, J_R$. They are related to the ordinary electric current $J$ and the chiral current $J_5$ as $J = J_L + J_R, J_5 = J_R-J_L$.
Anomaly results in the non - conservation of these currents.
 Following \cite{chiral_torsion2} we represent anomaly as a result of the spectral flow that follows from the explicit solution of Pauli equation. We extend the consideration of \cite{chiral_torsion2} to the case, when both emergent vielbein and emergent gauge field are present, and restrict ourselves to the case, when the emergent vielbein does not depend on time while $e_0^i = 0$ for $i = 1,2,3$. The  expressions for the anomalies in the quasiparticle currents have the form:
\begin{eqnarray}
\langle \partial_\mu J_{R\pm}^\mu \rangle &=&  \frac{1}{8 \pi^2} \,{\bf i}^{\mu\nu\rho\sigma}  \, \partial_{\mu} (B_{\nu}-A^{\pm}_{\nu})\, \partial_{\rho}(B_{\sigma}-A^{\pm}_{\sigma}), \nonumber\\ \langle \partial_\mu J_{L\pm}^\mu \rangle & = & - \frac{1}{8 \pi^2} \,{\bf i}^{\mu\nu\rho\sigma}  \, \partial_{\mu} (B_{\nu}-A^{\pm}_{\nu})\, \partial_{\rho}(B_{\sigma}-A^{\pm}_{\sigma}) \label{AaLR}
\end{eqnarray}
Here $+$ stands for the quasiparticle current at ${\bf K}^{(0)}$ while $-$ stands for that of $-{\bf K}^{(0)}$.
This expression does not contain the Nieh - Yan term $\frac{\Lambda^2}{8 \pi^2} \,{\bf i}^{\mu\nu\rho\sigma}  \, \partial_{\mu} e^{a}_{\nu}\, \partial_{\rho}e^b_{\sigma}\eta_{ab}$ (that is expressed through the ultraviolet cutoff\footnote{Very roughly it denotes the scale, at which the quasiparticle dispersion loses its Dirac cone form. That means, that $\Lambda \leq |{\bf K}|$. Here $\bf K$ is the position of the Dirac point, that obeys $|{\bf K}| \sim 1/a$, where $a$ is the interatomic distance.} $\Lambda$) simply because this term vanishes when the vielbein does not depend on time and $e_0^i = 0$ for $i=1,2,3$. We do not discuss the possible appearance of this term in the case of the time - depending vielbein.

Eq. (\ref{AaLR}) may be derived in the absence of torsion in a number of different ways (see, for example, \cite{Gorbar:2013dha,chiral_torsion2,Zyuzin:2012tv,tewary}). At the dislocation the emergent magnetic field is of the order of $\frac{1}{S_{}}$, where $S_{}$ is the area of the dislocation core. Torsion magnetic field is of the order of $\frac{1}{S_{}\Lambda}$ and is suppressed at low energies compared to the emergent magnetic field due to the cutoff $\Lambda$ in the denominator. Therefore, torsion may be considered as a perturbation. Moreover, in the absence of torsion electric field $T_{0i}^a$ it can be shown, that this perturbation due to torsion that appears due to the dislocation does not affect the spectral flow and the total expression for the chiral anomaly of Eq. (\ref{AaLR}) (at least, for the non - marginal cases, when the emergent magnetic field does not vanish and therefore dominates over the torsion magnetic field). In Appendix B we illustrate this pattern using the direct solutions of Pauli equation in the presence of the torsion magnetic field of particular form - when it is directed along the emergent magnetic field. This is the case realized in the presence of the dislocation.

In the marginal case ${\bf b} {\bf K} = 0$ the emergent magnetic field carried by the dislocation vanishes, and torsion cannot be considered as perturbation (see Sect. \ref{sectmarg}) of Appendix B). We do not work out this case in details and do not exclude, that in this case the emergent relativistic invariance may be broken by the anomaly.

Since the emergent $U(1)$ fields at the two Dirac points have opposite charges, and its time derivative vanishes, we arrive at the total chiral anomaly ($J_5 = \sum_{\pm}(J_{R,\pm}^\mu-J_{L,\pm}^\mu)$):
\begin{eqnarray}
\langle \partial_\mu J_{5}^\mu\rangle &=& \frac{1}{2 \pi^2} \,{\bf i}^{\mu\nu\rho\sigma}  \, \partial_{\mu} B_{\nu}\, \partial_{\rho}B_{\sigma}   \label{AaLR2}
\end{eqnarray}
It is given by the external electromagnetic field $B$ only.
The emergent magnetic field contributes the expression for the anomalous transfer of chirality between the two Dirac points:
\begin{eqnarray}
\langle \partial_\mu (J_{5,+}^\mu-J_{5,-}^\mu)\rangle &=& -\frac{1}{ \pi^2} \,{\bf i}^{\mu\nu\rho\sigma}  \, \partial_{\mu} B_{\nu}\, \partial_{\rho}A_{\sigma}   \label{AaLR3}
\end{eqnarray}
(Here we substitute $A=A^+$).

\subsection{Anomalous production of quasiparticles at the dislocation}

Now we are in the position to calculate the anomalous production of quasiparticles at the dislocation in the presence of external electric field. We suppose for simplicity, that the external magnetic field is absent. Therefore, the mentioned above marginal case ${\bf K} {\bf b}=0$ is trivial: there is no production of quasiparticles at all. At  ${\bf K} {\bf b}\ne 0$ we consider the gauge, in which ${\bf B} = 0, B_0 = \phi$. According to Eq. (\ref{AaLR}) the anomalous production of quasiparticles appears at each Dirac point given by
\begin{equation}
\langle \partial_\mu J_{R,\pm}^\mu\rangle \approx  \mp \frac{1}{4\pi^2}  \int_{l^0}   b^a {\bf K}_a \partial_{\mu} \phi  d {Y}^\mu(s) \delta^{(3)} (X-Y(s)), \quad \langle \partial_\mu J_{L,\pm}^\mu\rangle \approx  \pm \frac{1}{4\pi^2}  \int_{l^0}   b^a {\bf K}_a \partial_{\mu} \phi  d {Y}^\mu(s) \delta^{(3)} (X-Y(s))\label{dJ}
\end{equation}
Here $\pm$ refers to the Dirac point.
One can see, that the total chirality is conserved  $\langle \partial_\mu J_{}^\mu\rangle =\langle \partial_\mu J_{5}^\mu\rangle =0 $. At the same time there is the transition of chirality between the two Dirac points.

Let us denote the density along the dislocation of the total production rate of the left/right - handed quasiparticles (incident at the two Dirac points $\pm {\bf K}^{(0)}$)  by $\dot{q}^{\pm}_{L,R} = \frac{d}{d t}\frac{d}{dY} Q^{\pm}_{L,R}$, where $Q^{\pm}_{L,R}$ is the total number of the left-handed/right-handed quasiparticles with momenta around $\pm{\bf K}^{(0)}$.
We have:
\begin{eqnarray}
&& \dot{q}^{\pm}_R  \approx    \pm\frac{1}{4\pi^2 } \, \Big({\bf K}_j {\bf b}^j\Big) \, \Big( {\bf E}_k\hat{\bf n}^k\Big), \quad \dot{q}^{\pm}_L  \approx    \mp\frac{1}{4\pi^2 } \, \Big({\bf K}_j {\bf b}^j\Big) \, \Big( {\bf E}_k\hat{\bf n}^k \Big)\label{chp}
\end{eqnarray}
Here the unity vector along the dislocation is denoted by $\hat{\bf n}$ (we assume ${\bf n}^2=1$). This expression is valid also in the original reference frame (in which the Fermi velocity is anisotropic).

\subsection{Chiral magnetic effect along the dislocation}

The production of quasiparticles of Eq. (\ref{chp}) in the presence of external electric field results in the appearance of the chiral imbalance at each Dirac point. Let us suppose for definiteness, that the dislocation is a straight line connecting boundaries while the sample area in the plane orthogonal to the dislocation is $S_\bot$.

Suppose, that the considered above process of the quasiparticle pumping from vacuum takes place without external magnetic field $H_B$ due to the emergent magnetic field $H_A$ incident at the dislocation.  The created particles and holes leave the dislocation and take part in the scattering processes within the whole volume of the semimetal.
Following \cite{CME} we consider the simplified equation that describes the kinetic of the appearance of chiral imbalance at the Dirac point $\pm{\bf K}$ in the presence of external electric field $E$ directed along the dislocation:
\begin{equation}
\dot{\rho}^\pm_5 = \pm \frac{1}{4\pi^2 S_\bot}{\bf K} {\bf b} \, E - \frac{\rho^\pm_5}{\tau_V}
\end{equation}
Here the chirality changing scattering time is denoted by $\tau_V$, while the chiral charge density (averaged over the whole volume of the sample) is $2\rho_5^\pm = \rho_{R}^\pm - \rho_{L}^\pm $, while $\rho_{L,R}^\pm$ is the density of the left-handed/right - handed particles at the Dirac point $\pm {\bf K}$. We neglect scattering processes with the possible transitions between the fermions incident at the different Dirac points.
The solution of this equation for $\tau \gg \tau_V$ gives
\begin{equation}
{\rho}^\pm_5 \approx  \pm \frac{1}{4\pi^2 S_\bot}{\bf K} {\bf b} \, E {\tau_V}
\end{equation}
According to \cite{Fukushima:2008xe} in the presence of temperature $T$ and ordinary chemical potential $\mu$ the induced chiral chemical potential is related to $\rho^\pm_5$ as follows:
\begin{equation}
\rho^\pm_5 v_F^3 = \frac{1}{3 \pi^2} (\mu^\pm_5)^3 + \frac{1}{3}\mu^\pm_5 (T^2 + \frac{\mu^2}{\pi^2})
\end{equation}
(The thermalization occurs mostly out of the dislocation. Therefore, Eq. (43) of \cite{Fukushima:2008xe} can always be applied.)
For $\mu, T \ll \mu_5$ this results in $\mu^\pm_5 = v_F \Big(3 \pi^2 \rho_5\Big)^{1/3}$. However, following \cite{CME} we imply the more realistic situation, when $\mu, T \gg \mu_5$, and  get
\begin{equation}
\mu^\pm_5 = \frac{3 \rho^\pm_5 v_F^3}{T^2 + \frac{\mu^2}{\pi^2}}
\end{equation}
The quasiparticle currents along the dislocation are given by
\begin{eqnarray}
j_{R,\pm}^k &=&\int_0^{\mu^\pm_5} \frac{dp_z}{2\pi}\frac{\nu(p_3)}{ S_\bot}  \langle {\Psi}_{0,p_y,p_z}|e\, e^k_a \sigma^a |\Psi_{0,p_y,p_z}\rangle = \pm \frac{\mu^\pm_5}{4\pi^2}\int_{l^0} \Big({\bf K}_j {\bf b}^j\Big)  dY^k(s) \delta^{(3)} (X-Y(s)) \\  j_{L,\pm}^k &=& - \int^0_{-\mu^\pm_5} \frac{dp_z}{2\pi}\frac{\nu(p_3)}{ S_\bot}  \langle {\Psi}_{0,p_y,p_z} |e\, e^k_a\bar{\sigma}^a |\Psi_{0,p_y,p_z}\rangle = \pm \frac{\mu^\pm_5}{4\pi^2 } \int_{l^0} \Big({\bf K}_j {\bf b}^j\Big)  dY^k(s) \delta^{(3)} (X-Y(s))\nonumber
\end{eqnarray}
We used here, that $\langle e\,e^{\mu}_a {\bf n}^a \rangle = {\bf n}^\mu \langle 1-\partial u\rangle = {\bf n}^\mu$ along the dislocation.
These currents are directed along the dislocation.  The net electric current $j_{\rm CME}=\sum_\pm (j_R+j_L)$ originated from the chiral magnetic effect is the sum of the currents at the both Dirac points $\pm {\bf K}$:
\begin{eqnarray}
j^k_{\rm CME} &=&   \frac{3\, v_F^3\Big({\bf K} {\bf b}\Big)^2\,E\, \tau_V}{ 4 S_\bot \pi^4(T^2 + \frac{\mu^2}{\pi^2})}\int_{l^0}   dY^k(s) \delta^{(3)} (X-Y(s))
\end{eqnarray}
The total electric current through the sample is directed along the dislocation and is given by $J = \sigma_{\rm Ohm} E + \sigma_{\rm CME} E$, where $\sigma_{\rm Ohm}$ is the ordinary Ohmic conductivity while the contribution to conductivity originated from the chiral magnetic effect is given by:
\begin{eqnarray}
 \sigma_{\rm CME} &=&   \frac{3\, v_F^3 \Big({\bf K} {\bf b}\Big)^2 \tau_V}{4 S_\bot \pi^4(T^2 + \frac{\mu^2}{\pi^2})} \label{CMEcond}
\end{eqnarray}

\section{Conclusions}

\label{SectConcl}

Let us summarize the results obtained in the present paper.

\begin{enumerate}

\item{}

Quasiparticles in Dirac semimetal in the presence of dislocations experience emergent magnetic field and  emergent vielbein. Those emergent fields are expressed linearly through elastic deformations. Tensors of material parameters $\beta,\gamma$ enter the corresponding expressions.

\item{}

When the material parameters may be neglected, the dependence of emergent magnetic field and emergent vielbein on the elastic deformation in the presence of dislocations remains nontrivial. In this case the dependence is of pure geometrical origin. It results in the appearance of emergent magnetic flux and torsion singularity localized at the dislocation.

\item{}

The emergent magnetic flux and the torsion singularity carried by the dislocation may be observed using scattering of quasiparticles on the dislocation. Emergent magnetic flux results in the Aharonov - Bohm effect while the torsion singularity manifests itself in Stodolsky effect.

\item{}

In the presence of external electric field the quasiparticles are pumped from vacuum. This results in the appearance of chiral anomaly at each Dirac point and at the transition of chirality between the two Dirac points.

\item{}
We calculate the anomaly based production of quasiparticles along the dislocation in the presence of external electric field.

\item{}
The production of quasiparticles along the dislocation results in chiral imbalance. It drives the chiral magnetic effect, that is the appearance of electric current along the dislocation without any external magnetic field due to the emergent magnetic field incident at the dislocation.

\end{enumerate}

The given results confirm that the low energy effective theory of quasiparticles in Dirac semimetal (including the chiral anomaly) simulates relativistic high energy theory when the Coulomb interaction between the quasiparticles is neglected. Although the latter interaction is strong due to the large value of effective coupling constant \cite{sarma}, the experience of the investigation of graphene \cite{Katsbook} indicates, that for the consideration of various emergent relativistic effects it may be possible to neglect the interactions between the quasiparticles.

We observe, that the electric charge of the quasiparticles in the considered effective theory is conserved, while the chirality at each Dirac point is not. Chiral anomaly drives the production of quasiparticles at each Dirac point in the presence of parallel magnetic and electric fields. The dislocations give an effective gravitational background for the given low energy theory and the emergent magnetic flux located at the dislocation. In this particular case the gravitational background may be considered as perturbation. The chiral anomaly is dominated by the contribution of emergent magnetic field and results in the production of quasiparticles at the dislocations in the presence of external electric field without external magnetic field.

Moreover, the analogue of the chiral magnetic effect exists: there is the electric current directed along the dislocation (that is along the emergent magnetic field inside the dislocation). This current is proportional to the chiral chemical potential that appears due to the mentioned above pumping of the quasiparticles at the dislocation\footnote{Ordinary chiral magnetic effect in Dirac semimetals caused by the external rather than emergent magnetic field was discussed recently in \cite{CME}.}.

The discussed emergent relativistic effects in Dirac semimetals related to emergent gravity and emergent gauge field should affect transport phenomena and, therefore, may, in principle, be measured experimentally (see also \cite{Gorbar:2013dha,Son:2012bg,anomaly_semimetal}). Above we already mentioned one of the transport phenomena that is intimately connected with the emergent geometry:   the appearance of the contribution to conductivity related to the chiral magnetic effect, driven by the emergent magnetic field incident at the dislocation.

The investigation of Dirac semimetal in the presence of the dislocations opens the possibility to observe in laboratory the relativistic quantum field theory effects that may take place in particle physics at the energies that are not accessible at the present moment. These are the effects of the hypothetical space - time torsion. One of such effects (the Stodolsky effect) was discussed in the present paper.

In the present paper we did not considered the case of the time - depending strain that would result in the time depending vielbein. The appearance of the Nieh - Yan term in chiral anomaly in the Dirac semimetal (and the related observable effects) in this case remains an open question.
It would also be important to investigate the influence on the discussed phenomena of Coulomb interaction between the quasiparticles. Since this interaction is strong \cite{sarma}, most likely, such a study should include numerical lattice methods. Finally, we would like to notice, that the constructions discussed in the present paper may be extended to the consideration of Weyl semimetals, which, presumably, have a more rich phenomenology.

The author is greatful to G.E.Volovik and V.A.Miransky for useful discussions.
The work is  supported by the Natural Sciences and Engineering
Research Council of
Canada and by grant RFBR 14-02-01261.

\section*{Appendix A}

Here we give the solution of elasticity equation Eq. (\ref{elaste}) for the case, when the origin of the dislocation is the infinite straight line (\cite{Landau}, chapter 27, problem 4). We imply the decomposition of Eq. (\ref{using}).

In the presence of the dislocation we obtain the
equation for the continuous part of $u^i$:
\begin{eqnarray}
0 &=& (\Delta \delta^{ik} + \frac{1}{1-2\sigma} \nabla^i \nabla^k)\Big(u^k_{\rm cont}-b^l {\bf i}^{kl} \frac{{\rm log}\,|X|}{2\pi}\Big)+ 2 b^l {\bf i}^{il} \delta^{(2)}(X),\nonumber\\ &&  \quad {\bf i}^{12} = - {\bf i}^{21}=1, \quad {\bf i}^{11}={\bf i}^{22} = 0, \quad \hat{X}^i = \frac{X^i}{|X|}, \label{elastU}
\end{eqnarray}
Here we place the origin of the dislocation at $X^i = 0$. We come to the following solution:
\begin{eqnarray}
 u^k_{\rm cont}(X) &=&  b^l {\bf i}^{kl} \frac{({\rm log} \frac{|X|}{2 R} +\gamma)}{2\pi}- \frac{1}{\Delta}( \delta^{ik} - \frac{1}{2(1-\sigma)}\frac{ \nabla^i \nabla^k}{\Delta})   2 b^l {\bf i}^{il} \delta^{(2)}(X) \label{ucontsol0}  \\
&=& - \frac{b^l {\bf i}^{kl}}{2\pi}({\rm log} \frac{|X|}{2 R} +\gamma) +  \frac{1}{1-\sigma} \nabla^i \nabla^k  \frac{ b^l {\bf i}^{il}}{8\pi}\, ({\rm log} \frac{|X|}{2 R} +\gamma-1)|X|^2\nonumber
\end{eqnarray}
The inverse Laplace operator $\Delta^{-1}$ is to be defined taking into account boundary conditions and the finite size of graphene sample. For the sample of linear size $R$ we come to the following expressions:
\begin{equation}
- \Big[\Delta^{-1}\Big]_X = \int \frac{d^2 K}{(2\pi)^2}\frac{e^{i K^a X^a}}{|K|^2}\approx  - \frac{1}{2\pi}\, ({\rm log} \frac{|X|}{2 R} +\gamma), \quad |X| \ll R
\end{equation}
($\gamma$ is the Euler constant), and
\begin{equation}
\Big[\Delta^{-2}\Big]_X = \int \frac{d^2 K}{(2\pi)^2}\frac{e^{i K^a X^a}}{|K|^4}\approx \frac{R^2}{4\pi} +  \frac{1}{8\pi}\, ({\rm log} \frac{|X|}{2 R} +\gamma-1)|X|^2, \quad |X| \ll R
\end{equation}
Up to an irrelevant constant we have
\begin{eqnarray}
 u^k_{\rm cont}(X) &=&  -\frac{1-2\sigma}{2(1-\sigma)}  b^l {\bf i}^{kl}   \frac{1}{2\pi}  {\rm log} \frac{|X|e^{\gamma}}{2 R}+  b^l {\bf i}^{il} \frac{1}{1-\sigma} \frac{\hat{X}^i \hat{X}^k}{4\pi} \label{ucontsol1A}
\end{eqnarray}

\section*{Appendix B}

\subsection{Pauli equation in the presence of the torsion magnetic field directed along the dislocation. }

In principle, torsion may be considered as a perturbation while the zero approximation corresponds to the solution of Pauli equation with flat vielbein in the presence of emergent (or real) magnetic field and external electric field. One can check that the perturbation due to torsion does not affect the spectral flow at least, when there is the emergent magnetic field and emergent torsion magnetic field that both are caused by dislocations (see below, Sect. \ref{sectpert}).   Below we illustrate this pattern by the solution of Pauli equation in the presence of torsion field $T_{ij}^a$ directed along the dislocation, i.e. when the dislocation is directed along the third axis, the only nonzero component of torsion is $T_{12}^a$.
This is the case, realized inside the dislocation, where both emergent magnetic field and emergent torsion are directed along the dislocation.
For the purpose of illustration we shall consider the further simplification with the only nonzero component of torsion $T_{12}^3$. The extension to the case, when the components $T_{12}^2$ and/or $T_{12}^1$ are nonzero is straightforward.

Thus we consider the case when constant emergent magnetic field $H^\pm_A$ and real magnetic fields $H_B$, as well as the electric field $E=E_B$ are directed along the third axis. The external field is considered in the gauge $B_0 = - z E$, while  the vielbein has the form:
\begin{equation}
e^{a}_{\mu} = \delta^a_{\mu} + \delta^{a3} \delta_{\mu 2} T_H X^1
\end{equation}
As it was mentioned above, the nonvanishing component of torsion is given by $T^3_{12} = T_H$. It corresponds to the so - called torsion magnetic field \cite{chiral_torsion2}.
Pauli equation for the left - handed/right - handed fermions receives the form (we denote $H=-H^\pm_A+H_B$) \cite{chiral_torsion2}:
\begin{equation}
i\partial_t \Psi_{L,R} = \hat{\cal H}^{(E)}_{L,R}\Psi_{L,R}
\end{equation}
The corresponding Hamiltonian has the form
\begin{equation}
\hat{\cal H}^{(E)}_{L,R} = -E z \mp \Big( \sigma^1 (-i\partial_1) + \sigma^2(p_2 - T_H X^1 (-i \partial_3 ) + H X^1) + \sigma^3 (-i \partial_3 )\Big)\label{pauli2}
\end{equation}
It coincides with the usual Hamiltonian for the particle in the presence of electric field $E$ and magnetic field $\tilde{H} = H -T_H p_3$. The spectrum of operator Eq. (\ref{pauli2}) $\hat{\cal H}^{(0)}_L$ for $E=0$ is given by:
\begin{eqnarray}
{\cal E}^{(L)}_{l,s_z,p_3} &=& \pm \sqrt{p_3^2 + \Big|H - T_H p_3 \Big|(2l+1) + 2(H- T_H p_3)  s_z}, \quad l = 1,2,..., \quad s_z = \pm 1/2\nonumber\\
{\cal E}^{(L)}_{0,s_z,p_3} &=& {\rm sign}\Big(T_H p_3 -H\Big) p_3, \quad s_z = \frac{1}{2}{\rm sign}\, \Big(T_H p_3 -H\Big)
\end{eqnarray}
The wave function of the lowest state has the form:
\begin{equation}
\Psi_{0,s_z,p_2,p_3,t}(x,y,z) = e^{i p_2 y + i p_3 z - \frac{1}{2}\Big|-T_H p_3+H\Big|\Big(x + \frac{p_2}{-T_H p_3 +H}\Big)^2}| s_z\rangle
\end{equation}
Here $| s_z\rangle$ is the eigenstate of $\frac{1}{2} \sigma^3$ corresponding to the eigenvalue $s_z$. We omit the normalization factor and denote $X^{\mu} = (X^0,X^1,X^2,X^3)=(t,x,y,z)$.
In the same way we find the spectrum of the Hamiltonian for the right - handed particles. It differs from that of the left - handed ones  by the expression for the lowest Landau level (LLL):
\begin{eqnarray}
{\cal E}^{(R)}_{0,s_z,p_3} &=& -{\rm sign}\Big(T_H p_3 -H\Big) p_3 , \quad s_z = \frac{1}{2}{\rm sign} \, \Big(T_Hp_3-H\Big)
\end{eqnarray}
The degeneracy of the level is $\nu(p_3) = \frac{1}{2\pi} \Big|T_H p_3 -H\Big| S_\bot$, where $S_\bot$ is the area of the system in the plane orthogonal to the magnetic field.  In the following by the subscript $\|$ we denote the component of the vector parallel to magnetic field.

\subsection{Evolution in time of the wave function}

Evolution in time of the wave function is given by:
\begin{equation}
\Psi^{L,R}(t,x,y,z) = e^{-i \hat{\cal H}^{(E)}_{L,R}  \, t } \Psi^{L,R}(0,x,y,z)\label{evolut}
\end{equation}
In general case this evolution includes transitions between the eigenstates of ${\cal H}^{(0)}_{L,R}$.
However, when the values of $p_3$ and $t$ are sufficiently small and we may neglect $T_H(p_3 + E t)$ compared to $H$, then the time evolution of the wave function for the LLL is given by
\begin{equation}
\Psi^{L,R}_{0,p_2,p_3}(t,x,y,z) \approx e^{iEtz \pm {\rm sign}(H-T_H p_3) i\int_0^t (p_3+Et)dt   +i p_2 y + i p_3 z - \frac{1}{2}|H-p_3T_H|\Big(x + \frac{p_2}{H-T_Hp_3}\Big)^2}| - \frac{1}{2}{\rm sign}(H-T_H p_3)\rangle\label{Psit2}
\end{equation}
It gives the solution of Pauli equation for sufficiently small $p_3$ and $t$. Notice, that the wave function of Eq. (\ref{Psit2}) is not the eigenstate of ${\cal H}^{(E)}_{L,R}$  given by Eq. (\ref{pauli2}). However, it corresponds to the definite values of momenta:
\begin{equation}
\hat{p}_y \Psi^{L,R}_{0,p_2,p_3}(t,x,y,z) = p_2 \Psi^{L,R}_{0,p_2,p_3}(t,x,y,z), \quad \hat{p}_z \Psi^{L,R}_{0,p_2,p_3}(t,x,y,z) = (p_3 + Et) \Psi^{L,R}_{0,p_2,p_3}(t,x,y,z)\label{Psit2p}
\end{equation}
Moreover, we have
\begin{equation}
\hat{\cal H}^{(0)}_{L,R} \Psi^{L,R}_{0,p_2,p_3}(t,x,y,z) =  \mp {\rm sign}(H-T_Hp_3)\,(p_3 + Et) \Psi^{L,R}_{0,p_2,p_3}(t,x,y,z)\label{Psit2H}
\end{equation}
where $\hat{\cal H}^{(0)}_{L,R}$ is the Hamiltonian for $E = 0$ given by
\begin{equation}
\hat{\cal H}^{(0)}_{L,R} = \mp\Big( \sigma^1 (-i\partial_1) + \sigma^2((-i\partial_2) - T_H X^1 (-i \partial_3 ) + H X^1) + \sigma^3 (-i \partial_3 )\Big)\label{pauli3}
\end{equation}

The mentioned above extension to the case of nonzero $T_{12}^2$ results in the substitution $T_H p_3 \rightarrow \sum_{a=2,3}T_H^ap_a$. We do not need to consider the case of nonzero $T_{12}^1$ because the x axis may always be chosen directed in such a way, that $T_{12}^1=0$.
One can see, that inclusion of nonzero $T_{12}^2$ does not affect the main property of the spectrum: the state of Eq. (\ref{Psit2}) with the given  value of $p_3$ carries the definite time - depending momentum $p_z(t) = p_3 + Et$ and the definite eigenvalue of the unperturbed Hamiltonian ${\cal E}(t) = \mp {\rm sign}(H)\,p_z(t)$ (for sufficiently small $p_3$ and $t$). For the values of momenta of the order of $H/T_H$, however, the evolution of the wave function in time is more complicated. It involves the transitions between different eigenstates of ${\cal H}^{(0)}$.


 \begin{figure}[h]
\centering
\includegraphics[width=6cm]{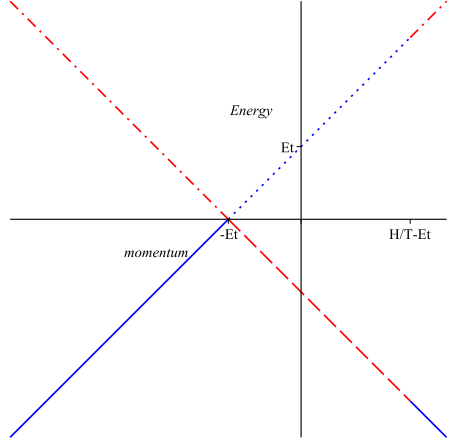}
\caption{\label{fig1}
The flow of the eigenvalues of ${\cal H}^{(0)}$ near the Dirac point in the presence of external electric field $E$, magnetic field $H$ and the torsion - magnetic field $T=T_H>0$. We represent schematically the dependence of the Hamiltonian eigenvalue ${\cal E}$ carried by the solution Eq. (\ref{Psit2})  on $p_3$ and on time. (Here $p_3$  is the difference between the third component of momentum  and the third component of the Dirac point). Crossing of the left - handed and the right - handed branches occurs at $p_3 = - E t$. At $p_3 = H/T_H - Et$ there is the discontinuity of the spectrum for each chirality. In vacuum the states with ${\cal E} < 0$ are occupied. Right - handed states are denoted by solid and dotted lines while the left - handed states are denoted by dashed and dashed - dotted lines. The value of $p_3$ belongs to the interval $[-\Lambda, +\Lambda]$, where $\Lambda \sim 2\pi/a$. Since $H_A\sim 2\pi/S_{\rm dislocation}$ while $T_H \sim a/S_{\rm dislocation}$, the discontinuities of the spectra do not appear within the Brillouin zone if $H_A\ne 0$, and may be neglected. }
\end{figure}

\subsection{LLL in the case ${\bf H}_A  \ne 0$}

 We consider the case, when the torsion electric field $T^a_{0i}$ vanishes while the torsion magnetic field $T_H$ is nonzero. We turn on electric field $E = E_B$ and magnetic fields $H = H_B - H_A$.
In the case of interest related to the emergent magnetic field located at the dislocation the typical values of magnetic field are of the order of $2\pi/S_{\rm dislocation}$, where $S_{\rm dislocation}$ is the area of the dislocation core (except for the case, when ${\bf K}{\bf b}=0$ so that the emergent magnetic field is absent at all). At the same time $T_H \sim a/S_{\rm dislocation}$, where $a$ is the interatomic distance. Therefore, $H/T_H \sim 2\pi/a \sim \Lambda$, and the left - handed states correspond to the left - moving particles while the right - handed states correspond to the right - moving particles (or vice versa depending on the sign of $H$).
The given pattern is illustrated by Fig. \ref{fig1}.
 \begin{figure}[h]
\centering
\includegraphics[width=6cm]{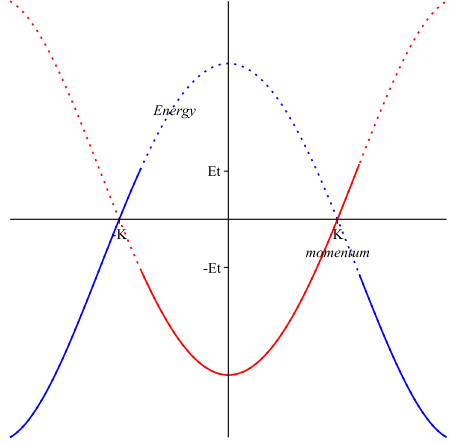}
\caption{\label{fig3}
Schematic pattern of LLL in the presence of external electric field $E$, magnetic field $H$ and the torsion - magnetic field $T_H>0$. We represent the eigenvalue ${\cal E}$ of ${\cal H}^{(0)}$ as a function of $p_3$, where $p_3$ is the third component of the total momentum (rather than the difference between the momentum and the position of the Dirac point). It is supposed, that the third axis is directed along ${\bf K}$. Crossing of the branches occurs at the two Dirac points  $\pm |{\bf K}|$. When the emergent magnetic field dominates, the left - handed branch at ${\bf K}$ is at the same time the left - handed branch at $-{\bf K}$. (The lines are closed through the boundary of the Brillouin zone.) The values of $\cal E$ carried by the occupied states at $t\ne 0$ are shifted by $Et$ (solid lines). As a result for $E\ne 0$ we observe the production of quasiparticles at each Dirac point. }
\end{figure}
The situation, when electric field is directed along the third axis of the Brillouin zone (then the coordinates of the Dirac points are $(0,0,\pm |{\bf K}^{(0)}|)$) is also illustrated by Fig. \ref{fig3}.

Both LLL branches of ${\cal H}^{(0)}_{L,R}$ are parametrized by two parameters $p_3$ and $p_2$. They have the meaning of momenta along the $z$ axis and $y$ axis only near the Dirac points, where, in addition, the dispersion ${\cal E}(p_2,p_3)$ does not depend on $p_2$. We can always organize the parametrization in such a way, that the line $p_2 = 0$ connects the two Dirac points $\pm{\bf K}^{(0)}$. For each value of $p_3$ we denote the number of states that belong to the given branch and are parametrized by $p_2$ by $\nu^{(R,L)}(p_3)$.
If the emergent magnetic field dominates, the LLL branch that represents the left - handed Weyl fermion located at ${\bf K}^{(0)}$ represents the left - handed Weyl fermion at $-{\bf K}^{(0)}$. The total number of states on the given branch for the values of $p_3$ varying between the positions of the zeros of $\cal E$ is equal to
 \begin{equation}
 N_{-} = \int \frac{L_\|dp_3}{2\pi} \nu^{(R)}(p_3)\, \theta\Big({\cal E}^{(R)}(0,p_3)\Big) , \quad  N_{+} = \int\frac{L_\|dp_3}{2\pi} \nu^{(L)}( p_3)\, \theta\Big({\cal E}^{(L)}(0,p_3)\Big)
 \end{equation}
 By $L_\|$ we denote the size of the system in the direction of magnetic field. Here ${\cal E}^{(R)}(p_2,p_3)$ and ${\cal E}^{(L)}(p_2,p_3)$ are the dispersions of the two branches of LLL that correspond to the right - handed and the left - handed particles at $\bf K$ correspondingly. The band structure in semimetal and our parametrization are organized in such a way, that the sign of ${\cal E}^{(R)}(p_2,p_3)$ (and of ${\cal E}^{(L)}(p_2,p_3)$) does not depend on $p_2$.
$N_-$ corresponds to the shortest path within the Brillouin zone, while $N_+$ corresponds to the path that goes through the boundary.  We denote by $\nu^{(R)}(p_3)$ the number of states with the given value of $p_3$ of the branch of spectrum that corresponds to the right - handed particles at ${\bf K}$. Correspondingly $\nu^{(L)}( p_3)$ is the number of states of the branch of spectrum that corresponds to the left - handed particles at ${\bf K}$.

\subsection{Dynamics of the spectral flow}

There are two distinct processes in the presence of electric field. The evolution in time of the states with the values of $p_3$ near the Dirac points is given by Eq. (\ref{Psit2}). This corresponds to the appearance of the quasiparticles/holes due to the spectral flow. The second process corresponds to the evolution in time of the states with the values of $p_3$ away from the Dirac points. Those states may transform to each other as it was mentioned above after Eq. (\ref{evolut}) if $T_H \ne 0$. Besides, the transitions may occur also within the other branches of spectrum and between the different branches. However, the latter transitions are suppressed by the difference in energy. Thus, we assume, that the states of LLL with $p_3$ between the positions of $-{\bf K}$ and  ${\bf K}$ remain occupied.

The state that corresponds to a certain value of $p_3$ near the Dirac point changes its energy as ${\cal E}(t,p_3) = {\cal E}(0,p_3+Et)$.
  As a result the system leaves its vacuum state (in vacuum all states below ${\cal E}=0$ are occupied). The production of the quasiparticles is given by
\begin{equation}
 \langle \partial J_{R} \rangle = \partial_t\int_{0}^{Et} \frac{d p_3}{2\pi} \frac{\nu^{(R)}(p_3)}{S_\bot} \approx  \frac{E\nu^{(R)}(0)}{2\pi S_\bot}     \approx    \frac{1}{4\pi^2} {H E}, \quad \langle \partial J_{L} \rangle \approx    - \frac{1}{4\pi^2} {H E}\label{AA}
\end{equation}
We see, that the presence of the torsion magnetic field $T_{12}^3$ does not affect the expression for the anomaly.
Our consideration may easily be extended to the case of nonvanishing components $T_{12}^1$ and $T_{12}^2$, that also do not contibute to anomaly.

We come to $\langle \partial J_{\pm} \rangle =0$ and
\begin{equation}
 \langle \partial J_{5,\pm} \rangle \approx  \frac{1}{2\pi^2} {(H_B-H^\pm_A)E}\label{J5HE}
\end{equation}
Eq. (\ref{AA}) may be easily generalized to the case of the slow varying fields with arbitrary components thus giving rise to Eq. (\ref{AaLR}).  We expect that this anomalous production of the quasiparticles should affect transport properties of semimetal (see also  \cite{Gorbar:2013dha,Son:2012bg,anomaly_semimetal}).

\subsection{Perturbations due to the arbitrary components of the vielbein}
\label{sectpert}

The extension to the case of nonzero $T_{13}^a$ and $T_{23}^a$ and nonzero  $\partial_{\{i}e^a_{j\}}$ (with $i,j = 1,2,3$) is more involved. The corresponding components of the vielbein may be considered as perturbations $\delta e^i_a = e^i_a - \delta^i_a$. The correction to the energy of the given state (for sufficiently small $p_3$) gives
\begin{equation}
 {\cal E}_{L,R} \approx \mp {\rm sign}(H) \, \langle{\bf d}\rangle {\bf p}
\end{equation}
 We denote ${\bf d} = (0,\delta e^2_3,1- \delta e^1_1 -  \delta e^2_2)$. Taking the leading terms in the expansion in the powers of momenta we may consider averages $\langle e^a_i \rangle$ over the state with $p_2=p_3=0$ as constants. Boundary between the states with positive energies and states with negative energies is the line that contains the point ${\bf p}=0$ and is orthogonal to $\langle{\bf d}\rangle$. If we neglect the transitions between the eigenstates of ${\cal H}^{(0)}$ with different values of $p_2,p_3$, then the spectral flow results in the shift of this boundary line towards the direction of the z - axis by $E t$. This gives the same production rate of quasiparticles as in the case of vanishing $\delta e^a_k$.

Notice, that the correction to the hamiltonian also results in the appearance of the transitions between the states with different values of $p_2,p_3$.  Matrix element between the states with  different values of ${\bf p}_{2,3}$ is
 \begin{equation}
\langle p^\prime|  \hat{\cal H} | p \rangle \approx \mp \frac{1}{2}{\rm sign}(H)  ({\bf p} +{\bf p}^\prime) \langle p^\prime|{\bf d}| p \rangle \mp \frac{i}{2} (p_2-p_2^\prime) \langle p^\prime| \delta e^1_3 | p \rangle \label{trans}
\end{equation}
As before, we neglect transitions between different Landau levels.
In principle, the transitions due to Eq. (\ref{trans}) complicate the consideration. However, both mentioned correction to energy and the transitions between the different eigenstates of the unperturbed hamiltonian are absent for the case of the emergent fields caused by the dislocation (because in this case $\delta e^a_3 = 0$).


\subsection{Marginal case $H/T_H \ll \Lambda$}
\label{sectmarg}

 As it was mentioned above, there exists the marginal case, when ${\bf b}{\bf K} = 0$ so that the emergent magnetic field carried by the dislocation vanishes. At the same time $T^a_H \sim b^a$, and we may have $H/T_H \ll \Lambda$, so that the pattern discussed above is changed. (Say, for $H=0$ all states in vacuum are left - handed or right - handed depending on the sign of $T_H$). As a result, the production of the quasiparticles with low energies near the Dirac points given by Eq. (\ref{AA}) should be followed by the additional appearance of the electrons/holes with energies $\sim H/T_H$. The corresponding states would experience the jump of energy $\sim  H/T_H$. This indicates, that the given process likely cannot be considered as adiabatic, and the whole consideration fails. For this reason, we do not consider the given marginal case in the present paper.

 \end{document}